\documentclass[aps,pra,showpacs,twocolumn,superscriptaddress]{revtex4}
\usepackage{times}
\usepackage{tabularx}
\usepackage{color}
\usepackage{amsmath,graphicx}
\newcommand{\ud}{\mathrm{d}}
\newcommand{\ic}{\mathrm{i}}

\def\ket#1{|#1\rangle}

\begin{document}

\title{Multifractality of quantum wave functions in the presence of perturbations}

\author{R. Dubertrand}
\affiliation{%
Universit\'e de Toulouse; UPS; Laboratoire de Physique Th\'eorique (IRSAMC); F-31062 Toulouse, France
}
\affiliation{CNRS; LPT (IRSAMC); UMR 5152, F-31062 Toulouse, France}
\affiliation{Institut de Physique Nucl\'eaire, Atomique et de
Spectroscopie, Universit\'e de Li\`ege, B\^at.\ B15, B - 4000
Li\`ege, Belgium}
\author{I. Garc\'ia-Mata}
\affiliation{Instituto de Investigaciones F\'isicas de Mar del Plata
(IFIMAR), CONICET--UNMdP,
Funes 3350, B7602AYL
Mar del Plata, Argentina.}
\affiliation{Consejo Nacional de Investigaciones Cient\'ificas y
Tecnol\'ogicas (CONICET), Argentina}
\author{B. Georgeot} 
\affiliation{%
Universit\'e de Toulouse; UPS; Laboratoire de
 Physique Th\'eorique (IRSAMC); F-31062 Toulouse, France
}
\affiliation{CNRS; LPT (IRSAMC); UMR 5152, F-31062 Toulouse, France}
\author{O. Giraud}
\affiliation{LPTMS, CNRS and Universit\'e Paris-Sud, UMR 8626, B\^at. 100,
91405 Orsay, France}
\author{G. Lemari\'e}
\affiliation{%
Universit\'e de Toulouse; UPS; Laboratoire de
 Physique Th\'eorique (IRSAMC); F-31062 Toulouse, France
}
\affiliation{CNRS; LPT (IRSAMC); UMR 5152, F-31062 Toulouse, France}
\author{J. Martin}
\affiliation{Institut de Physique Nucl\'eaire, Atomique et de
Spectroscopie, Universit\'e de Li\`ege, B\^at.\ B15, B - 4000
Li\`ege, Belgium}

\date{June 18, 2015}

\begin{abstract}
We present a comprehensive study of the destruction of quantum multifractality in the presence of perturbations. We study diverse representative models displaying multifractality, including a pseudointegrable system, the Anderson model and a random matrix model. We apply several types of natural perturbations which can be relevant for experimental implementations. We construct an analytical theory for certain cases, and perform extensive large-scale numerical simulations in other cases.  The data are analyzed through refined methods including double scaling analysis. Our results confirm the recent conjecture that multifractality breaks down following two scenarios. In the first one, multifractality is preserved unchanged below a certain characteristic length which decreases with perturbation strength. In the second one, multifractality is affected at all scales and disappears uniformly for a strong enough perturbation. Our refined analysis shows that subtle variants of these scenarios can be present in certain cases. This study could guide experimental implementations in order to observe quantum multifractality in real systems.

\end{abstract}
\maketitle

\section{Introduction}

Many physical systems display patterns that repeat themselves faithfully at every scale. When such systems are characterized by a single non integer dimension, they are called fractals (see e.g. \cite{mandelbrot}). More generically, multifractality corresponds to the case when different fractal dimensions are required to describe the system. Multifractality characterizes many complex classical phenomena: stock option analysis \cite{stock}, turbulence \cite{turbulence}, cloud imaging \cite{cloud}. 
In quantum physics a seminal example where multifractality occurs is the Anderson model for the transport of an electron in a disordered crystal \cite{Anderson58}. In the metallic phase the electron wave functions are spread uniformly inside the sample whereas in the insulator phase they are strongly localized. Exactly at the threshold of the transition wave functions show highly nontrivial fluctuations leading to anomalous transport. These fluctuations can be precisely described by a multifractal analysis, see e.g. \cite{mirlinRMP08} and references therein. Such types of multifractal wave functions can also be found in dynamical systems whose classical limit is neither integrable nor fully chaotic, which are dubbed pseudointegrable systems  \cite{Berry_pseudoint,bunimovitch_triangle_map,italo_prl}. Quantum multifractality in various related systems has been intensively studied on the theoretical side, from a condensed matter perspective \cite{kohmoto,mirlin2000,mirlinRMP08,romer,huckenstein,PRBM,PRBM2,pietronero,castellani,evangelou,manybody} for both one-body and many-body models and from a semiclassical point of view \cite{Giraud,coll,old,ossipov,indians1,indians,garciagarcia,interm,wiersig,MGG,eugene_charles_pseudoint,map_pseudoint,MGGG,BogGir}. However experimental characterization of multifractality has been much more challenging, despite some indirect recent attempts in disordered conductors \cite{richard} and cold atoms \cite{chabe2008experimental,lemarie2009observation,lemarie2010critical,lopez2013phase,cold2}. It is worth mentioning that a recent acoustics experiment simulating the Anderson model has allowed such a measurement \cite{billes}. 

As multifractality has been difficult to observe experimentally, it is crucial
to assess how it is affected by perturbations. This analysis is also important from a fundamental viewpoint, since  disturbances of the system may affect the wave function at different scales. Considering that multifractality is a multiscale phenomenon, this could lead to a wealth of possible behaviors.
The main goal of the present paper is thus to analyze how quantum systems with multifractal properties behave under the effect of an external perturbation. We have considered three paradigmatic one-body models, one being the Power law Random Banded Matrix model (PRBM), the second one the Anderson model, and the third one being representative of pseudointegrable systems. In these systems, we have investigated several natural perturbations in order to specify the robustness of quantum multifractality. At the same time these natural perturbations could account for real experimental situations. We have recently conjectured that quantum multifractality can be in general destroyed by a perturbation following two scenarios \cite{short}.  In scenario I, there exists a characteristic length below which multifractality is unchanged; the perturbation acts only by changing the characteristic length.  In scenario II,  multifractality is affected at all scales and vanishes uniformly when the perturbation increases. In the present paper, we confirm these two broad scenarios by new detailed analytical and numerical results.  We also introduce a double scaling analysis to describe a variant of the second scenario where a modified multifractality is observed only below a characteristic scale.  

In Sect.~\ref{defmodels} the models we have studied are more precisely introduced and the numerical methods used to obtain our results are described. In Sect.~\ref{multlocalperturb} we consider a first type of perturbation natural for pseudointegrable models, namely the smoothing of singularities in the potential. In Sect.~\ref{multfiniteN} we consider a change of parameters which moves the system away from criticality. In the case of a specific pseudointegrable system, we are able to predict the change of multifractality through an analytical theory that we expose in detail.  In Sect.~\ref{basis_change} we study the perturbation corresponding to a change of basis. Eventually we draw some conclusions in Sect.~\ref{conclusion}.

\section{Models and methods}
\label{defmodels}

\subsection{Models}
\label{defintermediate}

Many theoretical investigations on multifractals were first carried out 
on the example of the PRBM model \cite{PRBM} (see also \cite{mirlinRMP08,mirlin2000}). This model is defined (in the real periodic case) as the ensemble of symmetric $N\times N$ matrices with random real coefficients, with zero mean value, and a variance given, for $1\le i , j \le N$, by
\begin{equation}
  \label{PRBMvarcoeff}
  \left< H^2_{ii}\right>=1,\quad \left< H^2_{ij}\right>=\left[1+\left(\dfrac{N}{\pi b} \sin\left(\frac{\pi(i-j)}{N}\right) \right)^2 \right]^{-1}\ .
\end{equation}
The parameter $b$ (effective band width) allows to tune the multifractality of the model from a regime of strong multifractality ($b\ll 1$) to weak multifractality, where states are close to extended ($b\gg 1$).
We will use this model as a benchmark at specific places, especially since some analytical results are available \cite{mirlinRMP08,mirlin2000}.   However, it is not
related directly to physical models, and in most of the following we will concentrate on two other models of more immediate physical relevance. 

The second model we consider originates from semiclassical physics \cite{Giraud} (see also \cite{coll}) and describes the discrete time dynamics of one quantum particle kicked in one dimension with a classical limit between integrability and chaos (pseudointegrability). This model, called the intermediate map, is defined as the quantization of an interval-exchange map on the torus. The classical map is defined by
\begin{eqnarray}
  \label{classmap}
 p_{n+1}&=&p_n+\gamma \mod 1\ ,\nonumber\\
 x_{n+1}&=&x_n+2 p_{n+1} \mod 1\ .
\end{eqnarray}
It is generated by the following Hamiltonian, defined on the phase space as
\begin{equation}
  \label{defH}
  H(p,x)=p^2+V(x)\sum_n \delta(t-n)\ ,
\end{equation}
with $V(x)=-\gamma\{ x\}$,  where $\{x\}$ means the fractional part of $x$. 

For integrable systems motion in phase space is restricted to tori (surfaces of genus one), while for pseudo-integrable systems motion takes place on surfaces of higher genus. For the classical intermediate map with rational $\gamma=a/b$, motion with initial momentum $p_0$ is restricted to the $b$ one-dimensional tori (circles) $p=p_0+k\gamma$ with $0\leq k\leq b-1$, thus describing a surface of genus $b$. For irrational $\gamma$ the motion is ergodic as in chaotic systems, although no strong chaos is present.

The corresponding quantum map is a unitary operator $U$ on an $N-$dimensional Hilbert space. For the intermediate map it   is given in the momentum basis by the $N\times N$ unitary matrix
\begin{equation}
  \label{defU}
  U_{kl}
  =\frac{e^{-2\pi\ic k^2/N}}{N}\frac{1-e^{2\ic\pi\gamma N}}{1-e^{2\ic\pi(k-l+\gamma N)/N}}, \quad 0\le k,l \le N-1.
\end{equation}
 The dimension of the Hilbert space is related to the effective Planck constant $\hbar_{\rm eff}=2\pi/N$.  We also consider a random version of the model, where $e^{-2\pi\ic k^2/N}$ is replaced by $e^{-\ic \phi_k}$ \cite{eugene_charles_pseudoint}, with $\phi_k$ independent random variables uniformly distributed in $[0;2\pi]$. This model allows to get better statistics, and gives similar results as the non-random model, with some specificities that we will present.

The spectral statistics of the quantum map \eqref{defU} depend on the value of the parameter $\gamma$. For irrational $\gamma$, the spectral statistics follows the prediction for the Circular Unitary Ensemble (CUE) of random matrices characteristic of chaotic systems. For rational $\gamma=a/b$, the spectral statistics depend on the arithmetical properties of $b$ and are intermediate between the Poisson statistics of integrable systems and the random matrix result of  chaotic systems \cite{eugene_charles_pseudoint, map_pseudoint}. 
In the case where $\gamma$ is a rational number $\gamma=a/b$, the eigenvectors of the operator (\ref{defU}) in the momentum basis show multifractal properties \cite{MGGG}. The multifractality strength depends on $b$, from strong multifractality (small $b$) to weak multifractality (large $b$). 

The intermediate map corresponds to the quantization of a dynamical system. It is also known that multifractality can appear in the critical regime of disordered solid-state systems. To discuss this class of systems, we will consider the famous model proposed by Anderson in \cite{Anderson58}. The $d$-dimensional Anderson model is defined in the basis of lattice sites as
\begin{equation}\label{eq:HAnd}
 H = \sum_i \epsilon_i \vert i \rangle \langle i \vert + \sum_{\langle i, j\rangle} \vert i \rangle \langle j \vert \; , 
\end{equation}
where the random on-site energies $\epsilon_i$ are uniformly distributed in $[-W/2,W/2]$ and $\langle i, j\rangle$ denotes nearest neighbors. 
Eigenstates of this model (\ref{eq:HAnd}) are always exponentially localized in dimension one and two. The situation is different in three dimensions. Indeed for $d=3$ all eigenvectors are localized for large values of the disorder strength $W$, but the system performs a localization-delocalization transition at a value $W_c\approx 16.53$ \cite{romer}. For $W<W_c$, eigenstates in the vicinity of $E=0$ are extended. At the transition point $W=W_c$, states display multifractal properties \cite{mirlinRMP08}.

\subsection{Multifractal dimensions}
\label{num_methods}

\begin{figure}[!ht]
  \centering
  \includegraphics*[width=.45\textwidth]{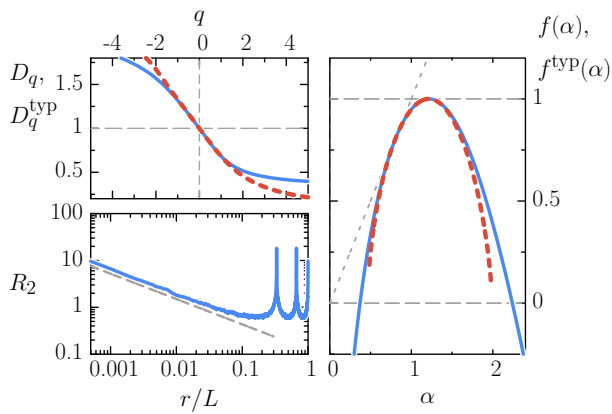}
  \caption{(Color online) Multifractal dimensions $D_q$ (top left) and singularity spectrum  $f(\alpha)$ (right) for the intermediate map for $\gamma=1/3$ and $N=2^{12}$ ; blue solid line: annealed exponent; red dashed line: typical.  Bottom left: correlation function $R_2(r)$; dashed gray line corresponds to the exponent $D_2-1$ obtained from top left.}
  \label{fig0}
\end{figure} 

There are several ways of defining multifractal dimensions for quantum states, which in most cases yield similar results \cite{MGGG}. In the present paper we will mainly use the box counting method. A system of linear size $L$ is decomposed into boxes of size $\ell$, and a coarse-grained measure of each box $k$ for a wave vector $\vert \psi \rangle$ is defined as $\mu_k=\sum_{i \in k} |\psi_{i}|^2$.
We define the moments $P_q(\ell)$ of order $q$ as
\begin{equation}
  \label{defPq}
P_{q}(\ell)=\sum_{k}\mu_k^q .
\end{equation}
In the limit of vanishing ratio $\ell/L$ between the box size and the system size, the presence of multifractality is characterized by the following behavior
\begin{equation}
   P_q(\ell)\sim \left(\frac{\ell}{L}  \right)^{\tau_q} , \qquad \ell/L\to 0\ , \label{limitPq}
\end{equation}
with a nontrivial exponent $\tau_q$.
The main quantity which will be used throughout this paper is the multifractal dimension $D_q\equiv \tau_q/(q-1)$.

 Another way of characterizing multifractality is to use the scaling of the moments as a function of the system size \cite{mirlin2000,mirlinRMP08}.  In the systems we study, this method has been shown to be equivalent to the box-counting method \cite{MGGG}. Here it will allow an analytical approach to be developed, which will be used in Section IV B. Nevertheless, this method can be delicate to use in certain cases, especially when looking at wave packets  \cite{MGGG12}. Another drawback is that in the systems we consider this approach makes it difficult to distinguish the physics at different scales, which is crucial in our study. 

Another signature of multifractality is the behavior of correlation functions such as the $2-$point correlation function
\begin{equation}
  \label{defR2}
  R_2(r)= N^2 \langle |\psi_i|^2 |\psi_{i+r}|^2 \rangle\ ,
\end{equation}
where $N=L^d$ is the Hilbert space dimension for a system of dimension $d$, and the average is taken over different eigenvectors, disorder (when present) and all indices $i$. It is related to the multifractal exponent $D_2$ \cite{mirlinRMP08,CatDeu87} via 
\begin{equation}
  \label{R2_D2}
  R_2(r)\sim  r^{D_2 -1} ,\qquad \frac{r}{L}\to 0.
\end{equation}

Alternatively, one may express multifractal properties via the singularity spectrum $f(\alpha)$, which is the Legendre transform of $\tau_q$. Moreover, for disordered systems, one distinguishes between the annealed exponents which describe the scaling of the average moments, and the typical exponents which characterize the average of the logarithm of the moments. For all the systems considered here, the two sets of exponents coincide over a relatively large range of $q$--values in the vicinity of $q=0$ \cite{MGGG}.  As an illustration, Fig.~\ref{fig0} shows an example of multifractal dimensions and singularity spectrum for the intermediate map.   In what follows we will mainly concentrate on the set of annealed exponents. We also display in Fig.~\ref{fig0} the correlation function $R_2$ for the intermediate map, together with the slope corresponding to $D_2$, illustrating relation (\ref{R2_D2}).

\subsection{Local multifractal exponents}
\label{scenarios}

Multifractality is mathematically defined as a scale invariance which takes place at all scales. In a real setting however, multifractality can be valid only on a certain limited range of scales, e.g.~between a lower microscopic length and an upper macroscopic length. As we will show, this is particularly relevant for perturbed systems. In order to investigate the ways in which multifractality is destroyed when a system is perturbed, one can introduce \cite{pietronero,short} a local multifractal exponent, which characterizes multifractality at a given scale. It is defined as
\begin{equation}
  \label{tildeDq}
  \tilde{D_q}(\ell)=\frac{1}{q-1}\frac{ \ud \ln P_q(\ell)}{\ud \ln \ell}.
\end{equation}
In practice as the scales $\ell$ are discrete numbers we compute the local multifractal exponent at scale $\ell$ as the slope between scale $\ell$ and the scale immediately above.

\begin{figure}[!b]
  \centering
  \includegraphics*[width=.45\textwidth]{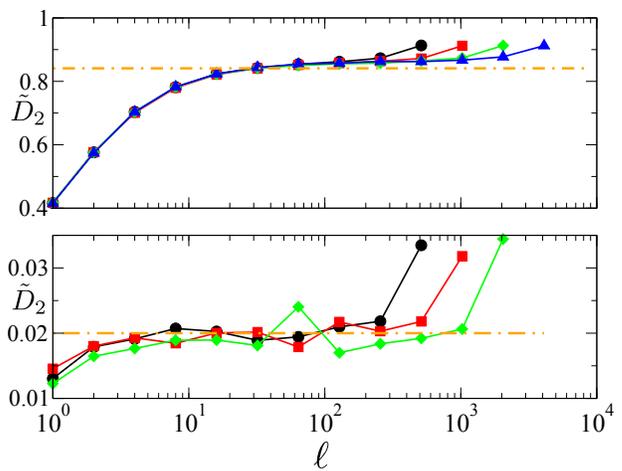}
  \caption{(Color online) Local multifractal dimension $\tilde{D}_2(\ell)$ as a function of the box size $\ell$ for the PRBM model \eqref{PRBMvarcoeff} for two values of $b$. Top: weak multifractality regime $b=2$. Bottom: strong multifractality regime $b=0.01$.  Black circles: $N=2^{10}$; red squares: $N=2^{11}$; green diamonds: $N=2^{12}$; blue triangles: $N=2^{13}$. 
Dotted dashed orange line: analytical prediction $D_2=2b$ for $b \ll 1$ and $D_2=1-\frac{1}{\pi b}$ for $b \gg 1$ \cite{mirlinRMP08}.}
  \label{Dqlocal_PRBMbis}
\end{figure} 

\begin{figure}[t!]
  \centering
  \includegraphics*[width=.45\textwidth]{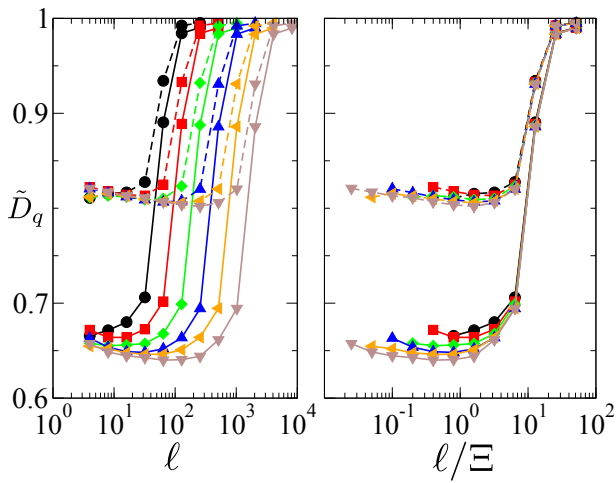}
  \caption{(Color online) Local multifractal dimensions for the intermediate map for $\gamma=1/5$ and increasing system sizes $N=L$. The full lines correspond to $q=2$ while the dashed lines correspond to $q=1$. Black circles: $N=2^9$; red squares: $N=2^{10}$; green diamonds: $N=2^{11}$; blue up triangles: $N=2^{12}$; orange left triangles: $N=2^{13}$; brown down triangles: $N=2^{14}$. Left: raw data. Right: all the box sizes are rescaled by $\Xi=N/5$.}
  \label{Dq_vsN}
\end{figure} 

The local multifractal exponents typically show a plateau which corresponds to the global multifractal exponent defined by (\ref{limitPq}); as an example, Fig.~\ref{Dqlocal_PRBMbis} displays the local 
multifractal exponents for the PRBM model, together with analytical predictions from \cite{mirlinRMP08}, and Fig.~\ref{Dq_vsN} displays this quantity for the intermediate map. Both figures show that  $\tilde{D_q}(\ell)$ indeed presents a plateau for a significant range of $\ell$ values. However, deviations can occur for the smallest values of $\ell$ as in Fig.~\ref{Dqlocal_PRBMbis}, which are due to the fact that coarse graining is necessary to obtain converged results. 

Deviations can also occur at large scales, as is the case for the intermediate map in Fig.~\ref{Dq_vsN}. The plateau at small $\ell$ coincides with the value $D_q$ from Fig.~\ref{fig0}, but at large scales $\tilde{D_q}(\ell)$ saturates to 1. This is a specificity of the model, which for $\gamma=a/b$ exhibits a characteristic length $\Xi\equiv N/b$, arising from the existence of the underlying classical structure described in section \ref{defintermediate}. The characteristic length can also be seen on the data shown on Fig.~\ref{fig0} for the correlation function, where $b$ peaks of typical width $N/b$ are clearly visible. Below the characteristic length $\Xi$, the value of  $\tilde{D_q}(\ell)$ shows a plateau indicating asymptotic multifractal behavior.
Figure \ref{Dq_vsN} shows that after rescaling the box size $\ell$ by this characteristic length, all $\tilde{D_q}(\ell)$ collapse to a single curve following the law 
 $\tilde{D}_q(\ell)=\mathcal{F}_q(\ell/\Xi)$, where $\mathcal{F}_q(u)$ is a function independent of $N$.


\subsection{Natural perturbations and scenarios}
\label{natperturb}

In the following Sections, several types of natural perturbations will be applied to these different models. For the intermediate map \eqref{defU} three types of perturbations naturally arise: i) the singular potential in \eqref{defU} can be smoothed; ii) the parameter $\gamma$ which controls multifractality can be varied away from its critical values; and iii) the measurement basis can be changed. Clearly, all these perturbations can arise in a real experimental setting, such as cold atom \cite{chabe2008experimental,lemarie2009observation,lemarie2010critical,lopez2013phase,cold2} or photonic lattice \cite{silberberg} implementations. In the case of the Anderson model \eqref{eq:HAnd}, the natural perturbations correspond to a change of disorder strength away from criticality and a change of measurement basis. 

 We will develop a scaling analysis of numerical data combined with analytical approaches in order to show that the different paths to multifractality breakdown always follow one of the two scenarios presented in \cite{short} and outlined in the introduction: scenario I corresponds to the existence of a characteristic length below which multifractality is unchanged; above this characteristic length which decreases with increasing perturbation strength, multifractality is destroyed. Scenario II corresponds to a multifractality which is affected at all scales and vanishes uniformly when the perturbation increases.  We will see that there can be interesting variations depending on the interplay between characteristic lengths of the model and the perturbations.

\section{Smoothing the singular potential}
\label{multlocalperturb}

In this section several types of smoothing of the intermediate map are described, which aim to account more realistically for experimental constraints. Indeed, in pseudo-integrable systems, generally singularities are present and are one of the reasons for which the classical dynamics is neither integrable nor chaotic. Experimentally, discontinuities such as in the potential in (\ref{defH}) have to be smoothed out.
In this section we will consider several types of smooth potentials approximating the exact one in different ways. These perturbations have been thought to be relevant for an experimental implementation of the intermediate map. One could envision photonic crystal implementations \cite{silberberg} where time is taken as a spatial dimension and the potential is etched on a substrate whose refractive index is varied. In this context, the potential singularity will be smoothed over a certain distance which depends on the etching technique. Another possible implementation corresponds to cold atom experiments where atoms are subjected to potentials constructed from laser light standing waves \cite{chabe2008experimental,lemarie2009observation,lemarie2010critical,lopez2013phase,cold2}. In this context, the smoothing of the potential will take place through the presence of only a fraction of the Fourier components needed to build the exact potential in  (\ref{defH}). Three possible ways of smoothing the potential $V(x)$ are considered below, adapted to these two experimental possibilities. We consider both the model with random phases and the deterministic model (\ref{defU}) (see section \ref{defintermediate}), more realistic for experiments.

\subsection{Polynomial smoothing}
\label{polynomialsmoothing}
\begin{figure}[!t]
    \centering  
    \includegraphics*[width=.45\textwidth]{fig4a.eps}\\[5pt]
     \includegraphics*[width=.45\textwidth]{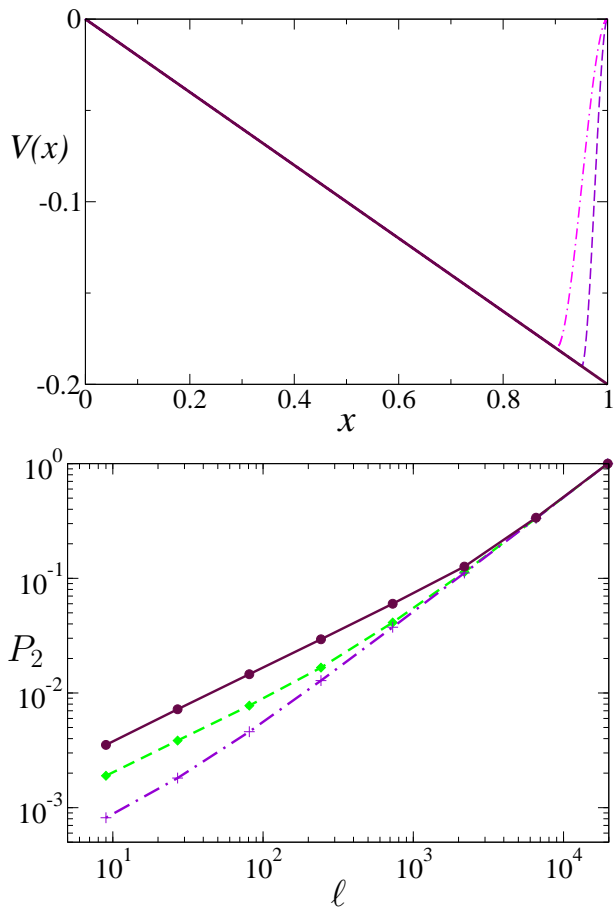}
   \caption{(Color online) Intermediate map smoothed by a polynomial for $N=3^{9}$ and $\gamma=1/5$. Top: Smoothed
potential. Dark brown full line: exact potential. Dashed purple line: $\epsilon=0.05$. Dot dashed magenta line: $\epsilon=0.1$.
Bottom: Moment of the eigenvectors of the random intermediate map for $q=2$ as a function of box size. Dark brown full line: exact potential; dashed green
line: $\epsilon=0.005$; dashed dotted purple line: $\epsilon=0.05$.
}
  \label{VsmoothP}
\end{figure}

\begin{figure}[!t]
  \centering
  \includegraphics*[width=.45\textwidth]{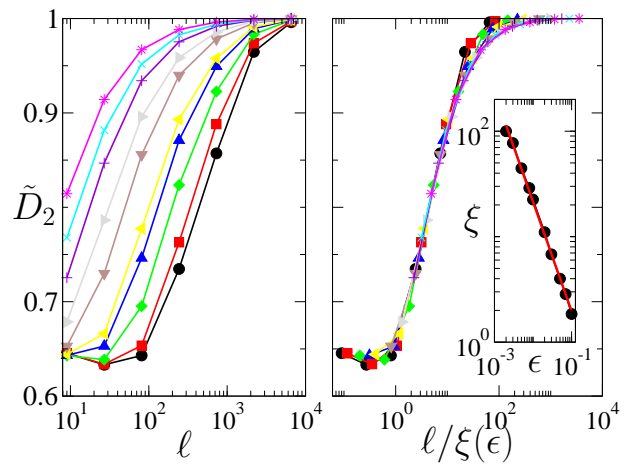}
  \caption{(Color online) Scaling analysis of the local multifractal dimension $\tilde{D}_2$ for the perturbed random intermediate map for $N=3^9$ and $\gamma=1/5$. The perturbation is the smoothing of the singular potential over a length $\epsilon$. Left: raw data. Right: data after rescaling. Inset: variation of the scaling length $\xi$ as a function of $\epsilon$; solid line is the fit corresponding to \eqref{scalDqtbis} with $\alpha = 1.04$. Black circles: $\epsilon=0.002$; red squares: $\epsilon=0.003$; green diamonds: $\epsilon=0.005$; blue up triangles: $\epsilon=0.008$; yellow left triangles: $\epsilon=0.01$; brown down triangles: $\epsilon=0.02$; gray right triangles: $\epsilon=0.03$; purple plus: $\epsilon=0.05$; cyan crosses: $\epsilon=0.07$; magenta stars: $\epsilon=0.1$. }
  \label{DqsmoothP}
\end{figure}

We first consider a more realistic version of the model for photonics experiments \cite{silberberg}.  In this context, 
we chose to approximate the potential $V(x)$ as
\begin{equation}
  \label{Vq_polynom}
    V(x)=\left\{
    \begin{array}{cl}
      -\gamma x,& 0\le x<1-\epsilon\\
      a_3 x^3+a_2 x^2+a_1 x+a_0,&1-\epsilon< x\le 1
    \end{array}\right. \ ,
\end{equation}
where the $a_i$ are chosen to make the potential and its first derivative continuous at $x=1-\epsilon$ and $x=1$. The original model (\ref{defU}) is recovered when $\epsilon=0$ so that $\epsilon$ can be seen as a small perturbative parameter.
Typical examples of the resulting potential are shown in Fig.~\ref{VsmoothP} top, while typical results for the moments $P_q(\ell)$ of the random intermediate map are shown in Fig.~\ref{VsmoothP} bottom for $N=3^9$. 
In Fig.~\ref{DqsmoothP} the local multifractal exponent $\tilde{D}_2(\ell)$ is plotted as a function of $\ell$ for several smoothing widths $\epsilon$. In the left panel the raw data are shown: while at very small perturbation strength $\epsilon$ one  observes a plateau at the unperturbed $D_2$ (compare with Fig.~\ref{Dq_vsN}), at larger values of $\epsilon$ this plateau is no longer visible and the curves $\tilde{D}_2(\ell)$ increase monotonically. Nevertheless, it turns out that one can put all these different curves onto a single one by rescaling the lengths $\ell$ (right panel). This shows that the local multifractal exponents obey a scaling relation:
\begin{equation}
  \label{scalDqt}
  \tilde{D_q}(\ell)=\mathcal{G}_q\left(\frac{\ell}{\xi(\epsilon)}\right), 
\end{equation}
with $\xi(\epsilon)$ a scaling length which depends only on the perturbation strength and which is well fitted by 
 \begin{equation}
  \label{scalDqtbis}
 \xi(\epsilon) \propto \epsilon^{-\alpha}
\end{equation}
with $\alpha \approx 1$ (see inset of Fig.~\ref{DqsmoothP}) and $\mathcal{G}_q$ is a scaling function independent of $\epsilon$.
This scaling behavior is valid for various values of the parameters $q$ and $N$. Indeed, it was shown in \cite{short} to occur when $N$ is of the form $2^n$ while Fig.~\ref{DqsmoothP} shows that it remains also valid for $N$ of the form $3^n$. Moreover, this scaling behavior with exponent $\alpha\approx 1$ applies for different values of $q$ and $\gamma$, and also for polynomial smoothings \eqref{Vq_polynom} of higher order (data not shown).
\begin{figure}[]
    \centering  
    \includegraphics*[width=.45\textwidth]{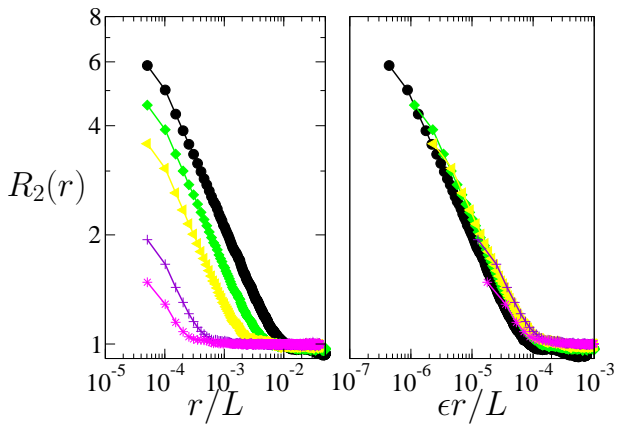}
  \caption{(Color online) Two-point correlation function $R_2$ as defined in (\ref{defR2}) for the random intermediate map smoothed by a polynomial for $N=3^{9}$ and $\gamma=1/5$.  Left: raw data. Right: data after rescaling the variable $r$ by $\xi =1/\epsilon$ following \eqref{scalDqtbis}. The same color code as in Fig.~\ref{DqsmoothP} is used.}
  \label{R2_smoothP}
\end{figure} 

Multifractality in this perturbed model can be further investigated using the $2-$point correlation function $R_2(r)$ as defined in (\ref{defR2}). In Fig.~\ref{R2_smoothP},  $R_2(r)$ is shown for several smoothing widths. Rescaling $r$ by the same scaling parameter $\xi(\epsilon)\propto 1/\epsilon$ leads to the collapse of all the curves, see Fig.~\ref{R2_smoothP}~right.

\begin{figure}[!t]
  \centering
  \includegraphics*[width=.45\textwidth]{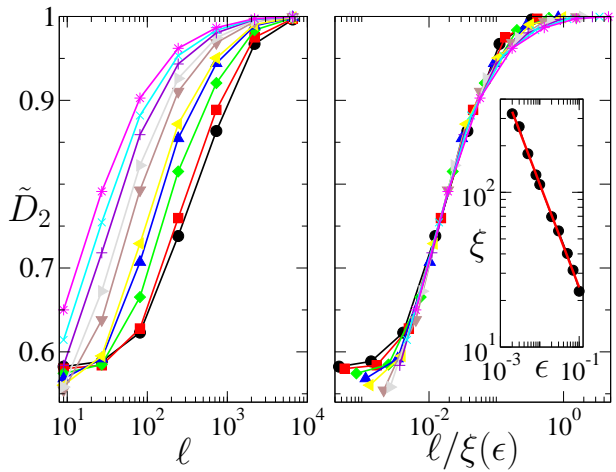}
  \caption{(Color online) Same as Fig.~\ref{DqsmoothP} for the non-random intermediate map  for $N=3^{9}$ and $\gamma=1/5$. Left: raw data of the local multifractal dimension $\tilde{D}_2$ for different perturbation strengths. Right: data after finite-size scaling. Inset: Variation of the scaling length $\xi$ as a function of $\epsilon$; solid line is the fit corresponding to \eqref{scalDqtbis} with $\alpha = 0.67$, different from the result in Fig.~\ref{DqsmoothP}. The same color code as in Fig.~\ref{DqsmoothP} is used.}
  \label{DqsmoothPnotrandom}
\end{figure}

A similar behavior can be observed for the model \eqref{defU} with non-random phases. In Fig.~\ref{DqsmoothPnotrandom} the local multifractal exponent $\tilde{D}_2(\ell)$ for the deterministic model is shown to follow the scaling law \eqref{scalDqt}, in complete analogy with Fig.~\ref{DqsmoothP}. 
\begin{figure}[]
    \centering  
    \includegraphics*[width=.45\textwidth]{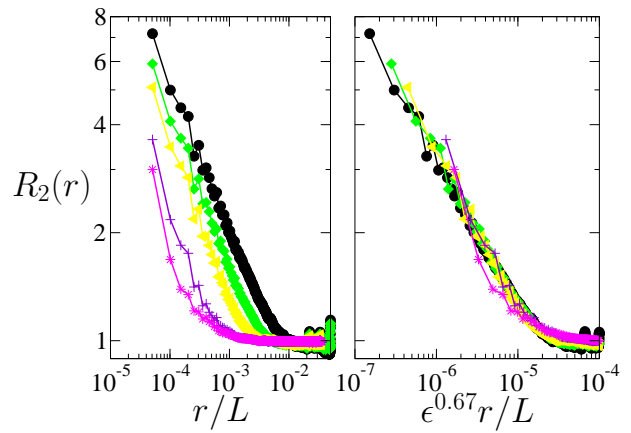}
  \caption{(Color online) Same as Fig.~\ref{R2_smoothP} for the non-random intermediate map  for $N=3^{9}$ and $\gamma=1/5$. Left: $2-$point correlation function as defined in (\ref{defR2}). Right: the variable $r$ is rescaled by $\xi(\epsilon) = \epsilon^{-0.67}$, different from the scaling used in Fig.~\ref{R2_smoothP}.
 The same color code as in Fig.~\ref{DqsmoothP} is used. }
  \label{R2_smoothPnotr}
\end{figure} 

The noticeable difference lies in the exponent $\alpha$ of the scaling length $\xi(\epsilon)$ with respect to the smoothing width $\epsilon$, which turns out to be $\alpha = 0.67$ rather than $\approx 1$ for the random phase model. As in the case of the random model, this value of  $\alpha\approx 0.67$ does not depend on $N$ or $\gamma$. Results for the $2-$point correlation function are shown in Fig.~\ref{R2_smoothPnotr} using the same parameters as in Fig.~\ref{R2_smoothP}. Again the same scaling behavior is observed, with an exponent $\alpha=0.67$ for the rescaling of $r$. The difference of the scaling exponent $\alpha$ between the random and non-random model reflect the differences of the correlations in the phases of the propagator coefficient (\ref{defU}).

 The data discussed in this section show that this kind of smoothing leads to the appearance of a characteristic length below which multifractality is unchanged, indicating that in this case multifractality breakdown occurs following scenario I of Subsection \ref{natperturb}.

\subsection{Fourier series smoothing}
In a cold atom experiment the potential experienced by the atoms can be created with standing waves of laser light. In these setups the frequencies and the amplitude can be controlled with a very high accuracy. One could think that each Fourier component of a periodic potential can then be simulated by one laser so that any potential could be reproduced. The limitation is that it is practically impossible to use a large number of lasers so that only potentials with a small number of non zero Fourier components can be modeled. 
The potential of the intermediate map is acting on the torus so it can be expanded as a discrete Fourier series. In this section we will investigate how the eigenvector statistics changes when the Fourier expansion of the potential is truncated to $N_f$ terms. 

\begin{figure}[b!]
  \centering
  \includegraphics*[width=.45\textwidth]{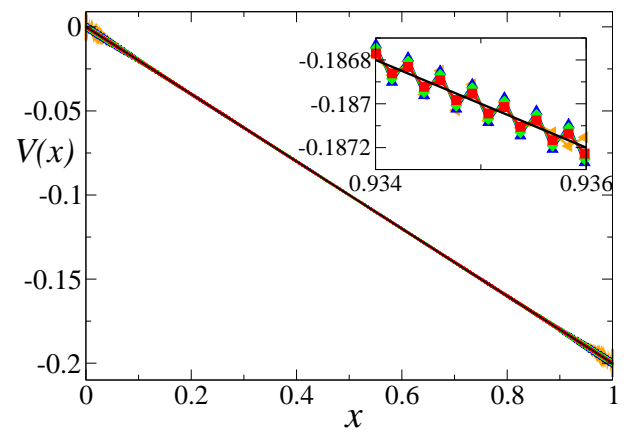}
  \caption{(Color online) Potential of the intermediate map obtained by truncation of Fourier series for $\gamma=1/5$ and $N=3^8=6561$ for different $N_f$.
Black full line: $N_f=N=6561$; red squares: $N_f=6559$; green diamonds: $N_f=6557$; blue up triangles: $N_f=6555$; orange left triangles: $N_f=6001$. Inset is a blow up of the same data.}
  \label{Vq_Nf}
\end{figure}
\begin{figure}[t!]
  \centering
  \includegraphics*[width=.45\textwidth]{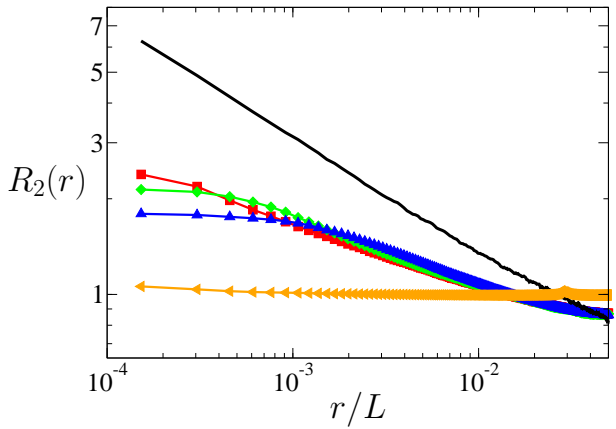}
  \caption{(Color online)  Two-point correlation function for the random intermediate map with potential approximated by a truncated Fourier series for $N=3^8$ and $\gamma=1/5$. Contrary to Fig.~\ref{R2_smoothP} we do not observe a systematic dependence on the truncation, indicating that the approximation is never perturbative.  The same color code is used as in Fig.~\ref{Vq_Nf}.
  }
  \label{R2trunc}
\end{figure}

 
For $N_f=N$ the linear form is recovered. A plot of the potential for different values of $N_f$ is shown in Fig.~\ref{Vq_Nf}. Contrary to the preceding case, the modification of the potential is not local anymore. In particular, even for large values of $N_f$ oscillations remain visible far from the discontinuity, see the inset in Fig.~\ref{Vq_Nf}. 
In this case our investigation shows that even for $N_f$ close to $N$, when almost all the Fourier components are kept, multifractality is completely destroyed. This indicates that this kind of perturbation is always large, and cannot be made arbitrarily small due to the discreteness of the Fourier series. This is illustrated by considering the $2$--point correlation function in Fig.~\ref{R2trunc}: contrary to Fig.~\ref{R2_smoothP} we do not observe a systematic dependence on the perturbation strength, even for $N_f$ close to $N$. Our results therefore show that a naive truncation of the Fourier series is not a good approach to experimentally observe multifractality in such systems. 

\subsection{Trigonometric smoothing}

\begin{figure}[t!]
  \centering
  \includegraphics*[width=.45\textwidth]{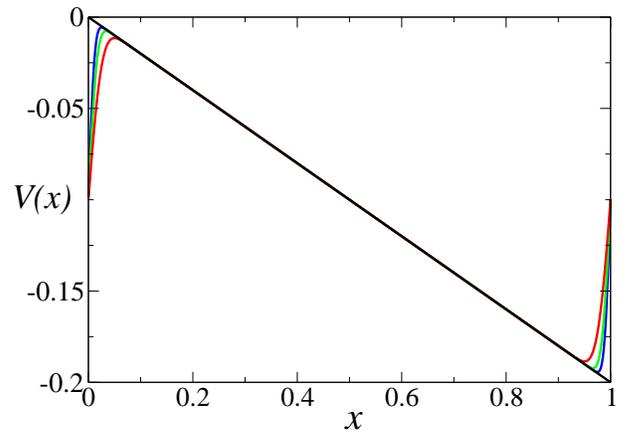}
  \caption{(Color online) Potential of the intermediate map approximated by a trigonometric smoothing with $N_d$ derivatives fixed at $x=0.5$, $N=2^{12}$ and $\gamma=1/5$. From right to left on the left: $N_d=46$ (red), $N_d=80$ (green), $N_d=109$ (blue); black straight line is the exact potential.}
  \label{Vq_Nd}
\end{figure}

In view of the results of the preceding subsection, one may try to search for better approximation schemes using a modified Fourier expansion. Indeed, a more efficient  way to approximate the potential for cold atom experiments can be devised by fixing a prescribed number $N_d$ of derivatives at one point in order to force the potential to be approximately linear around that point. Such a potential can be chosen as a trigonometric sum:
\begin{equation}
  V(x)=\sum_{l=0}^K a_l \sin(\pi l x)\ .
\end{equation}
We took $K=3N_d$ and the $a_l$ are fixed by the $N_d$ equations
\begin{eqnarray}
  V'(0.5)=-\gamma,\ V''(0.5)=0,\ , \dots,\ V^{(N_d)}(0.5)=0\ .
\end{eqnarray}
The resulting potential for several values of $N_d$ is shown in Fig.~\ref{Vq_Nd}.
\begin{figure}[]
  \centering
  \includegraphics*[width=.45\textwidth]{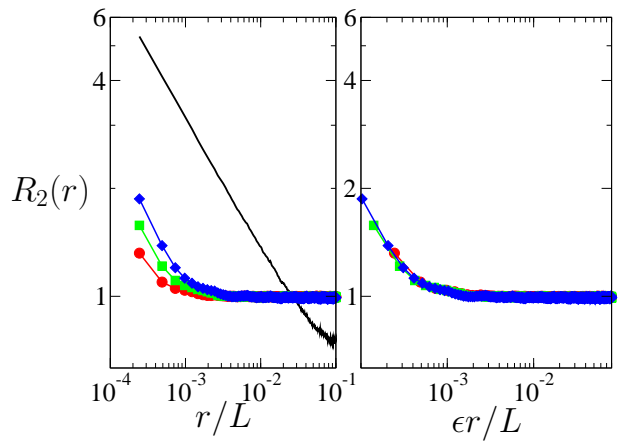}
  \caption{(Color online) Scaling analysis of the $2-$point correlation function for the random intermediate map with potential approximated by a trigonometric smoothing with $N_d$ derivatives fixed at $x=0.5$ for $N=2^{12}$ and $\gamma=1/5$. Black full line: exact model. Red circles: $N_d=46$. Green squares: $N_d=80$. Blue diamonds: $N_d=109$. Left: raw data; right: data rescaled with \eqref{scalingNd} with  $\epsilon=1/N_d$.}
  \label{R2_vs_Nd}
\end{figure} 
Compared with the simple truncation of the Fourier series, the potential change now occurs only in a limited region of space, which gets smaller and smaller as $N_d$ increases.
We found that a scaling analysis similar to the one in Subsection \ref{polynomialsmoothing} is possible in this case. As an example, the two-point correlation function for different $N_d$ is displayed in Fig.~\ref{R2_vs_Nd}. The scaling relation follows the formula
\begin{equation}
  \label{scalingNd}
  R_2(r)=\mathcal{R}\left(\frac{r}{\xi(\epsilon)}\right), \quad \xi(\epsilon) \propto \frac{1}{\epsilon}\ ,
\end{equation}
where $\epsilon=1/N_d$ and $\mathcal{R}$ is a scaling function independent of $N_d$. This is similar to what is described in Fig.~\ref{R2_smoothP}.

This way of expanding the potential as a trigonometric series is thus more efficient in order to keep the multifractality of the system, and the disappearance of scale invariance corresponds to scenario I (see Subsection \ref{natperturb}). 

\section{Moving a parameter away from criticality}
\label{multfiniteN}

In all the models considered, multifractality is predicted only for certain critical values of a parameter. In this Section, we
investigate the robustness of multifractal properties when this parameter is moved away from criticality.

\subsection{Change of $W$ in the Anderson model}

In the case of the 3D Anderson model, multifractality appears at the metal-insulator transition which corresponds to a specific disorder strength $W_c \approx 16.53$ \cite{romer}. A natural choice of perturbation is therefore to change the disorder strength slightly below or above the critical value $W_c$. In this case, it is known that the eigenstates are either localized or delocalized with a characteristic length $\xi$. In the insulating phase $\xi$ corresponds to the localization length, while in the metallic phase it corresponds to the correlation length. Below this characteristic length, the wave functions are multifractal with the same critical multifractal spectrum, and they form a ``multifractal insulator'' or a ``multifractal metal'' \cite{kravcuev}. This is a consequence of a one-parameter scaling law that governs the multifractal spectrum in the vicinity of the transition \cite{romer}. This type of perturbation therefore follows scenario I of quantum multifractality breakdown: quantum multifractality survives unchanged below a certain characteristic length related to the distance to the critical point.

\subsection{Change of $\gamma$ in the intermediate map}

\subsubsection{Numerical results}

\begin{figure}[ht]
  \begin{center}
\includegraphics*[width=0.42\textwidth]{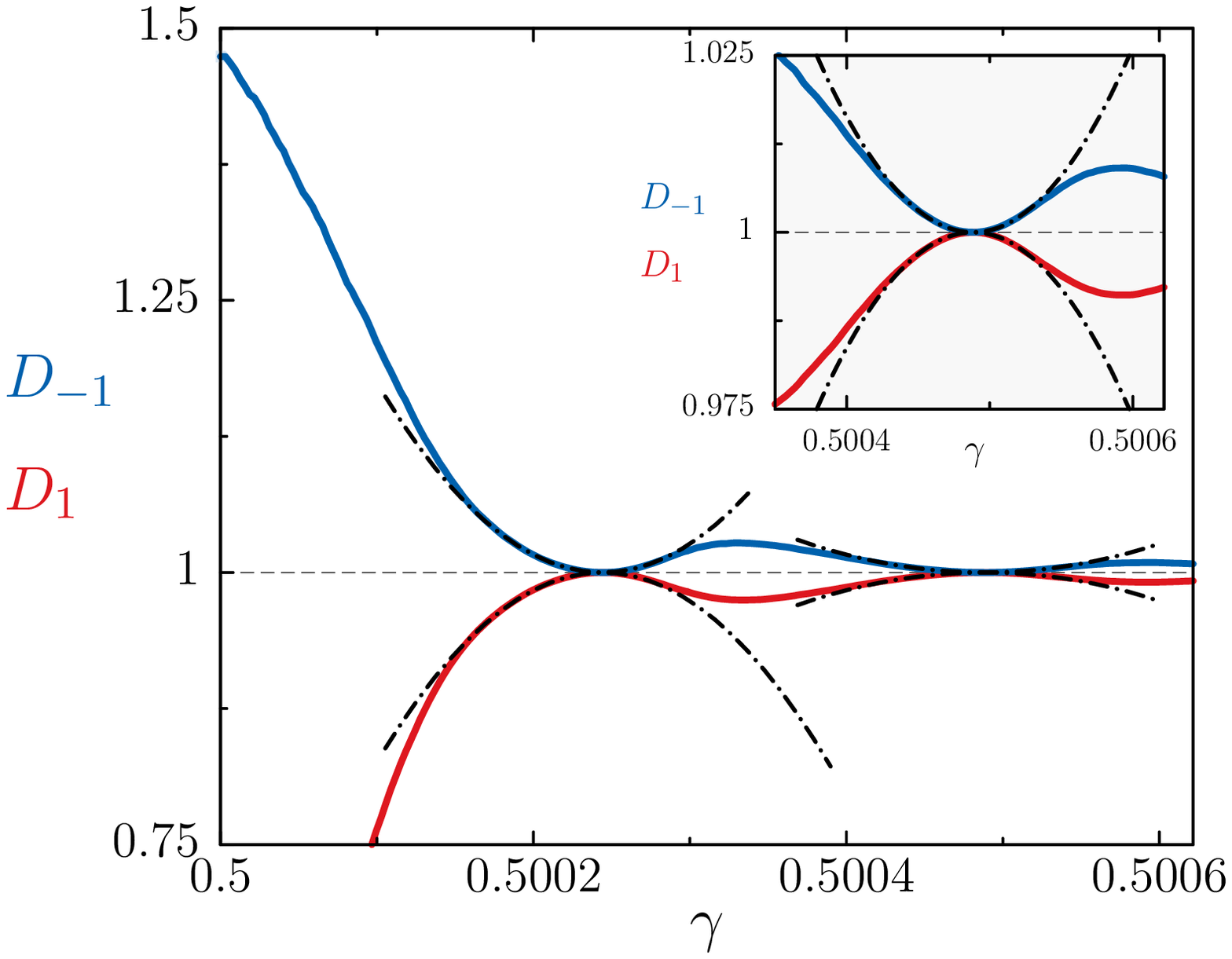}\\[10pt]
\includegraphics*[width=0.42\textwidth]{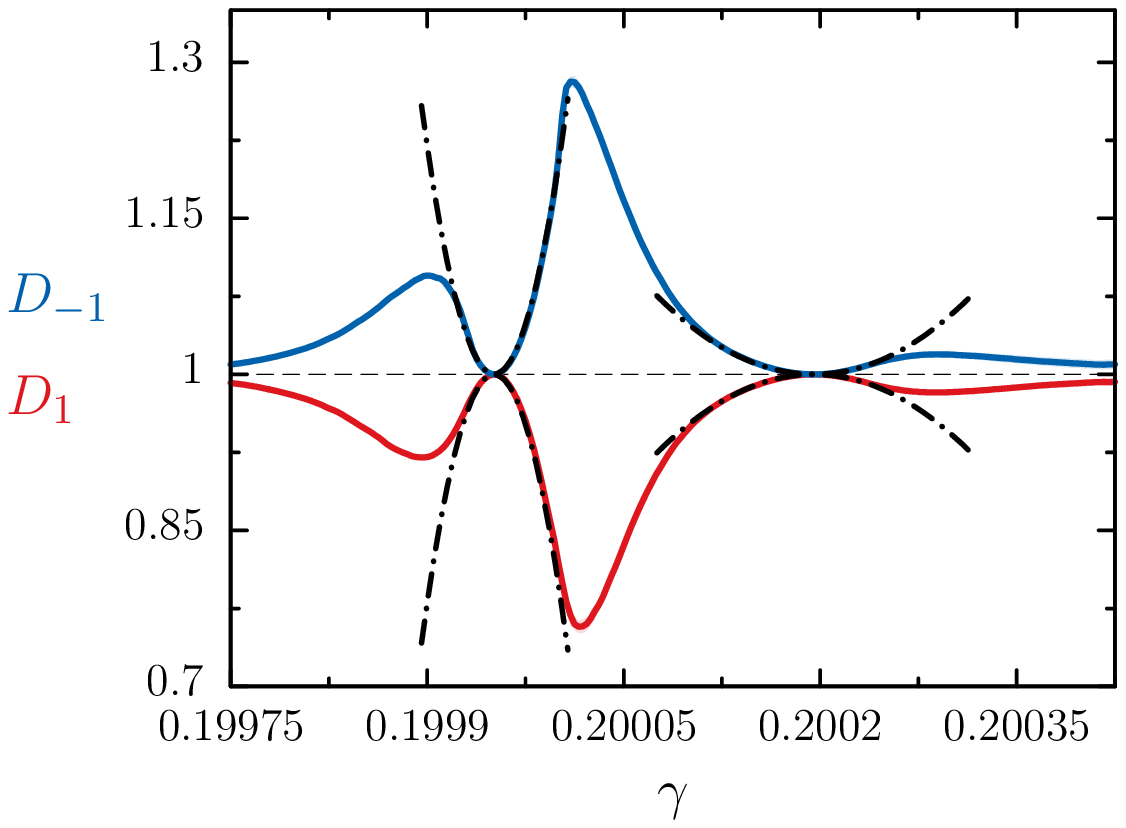}
\end{center}
\caption{(Color online) Variation of the multifractal dimensions $D_{\pm 1}(\gamma)$ for the intermediate map \eqref{defU} with random phases and
$N=2^{12}$ in the vicinity of the rational values $\gamma = 1/2$ (top) and  $\gamma = 1/5$ (bottom). The red solid line below 1 corresponds to $D_{1}(\gamma)$, the blue one above 1 to $D_{-1}(\gamma)$. Black dash-dotted parabolas correspond to the theoretical expression Eq.~\eqref{dqfinalrandom} with $\kappa=1$ and $2$ for $\gamma = 1/2$ and $\kappa = 0$ and $1$ for $\gamma = 1/5$. Inset: zoom on the right part of the top plot, corresponding to $\kappa=2$.
\label{Dq_vsg}}
\end{figure}

The behavior of the intermediate map is richer, in the sense that there is in principle an infinite number of critical values of the parameter, i. e. values for which multifractality arises. Indeed, in the intermediate map (\ref{defU}), multifractality is predicted to appear for rational values of the parameter $\gamma$ (see Section \ref{defmodels}). Multifractality manifests itself all the more strongly when the denominator of $\gamma$ is small, that is, for $\gamma=1/2,1/3$ or $1/5$ for instance \cite{MGG}. In close vicinity of these rationals, one should observe delocalized eigenstates. However, since this difference in behavior only arises in the limit of infinite size, we can expect at finite size $N$ a persistence of multifractal properties if one varies the parameter in some vicinity of these low-denominator rationals. This is illustrated in Fig.~\ref{Dq_vsg}, where $D_q$ is computed for a fixed vector size $N$ as a function of the parameter $\gamma$ in the vicinity of two rationals, $1/2$ and $1/5$, up to a distance of the order $1/N$ from these rationals. Clearly the curve $D_q(\gamma)$ is not singular at all rationals, but rather smoothed out. An advantage of this model is that it is amenable to analytical treatment via perturbation theory, which allows us to get a clear picture of how parameter changes may affect multifractality. This approach will be carried out in the next subsection.

\begin{figure}[t]
\includegraphics*[width=0.45\textwidth]{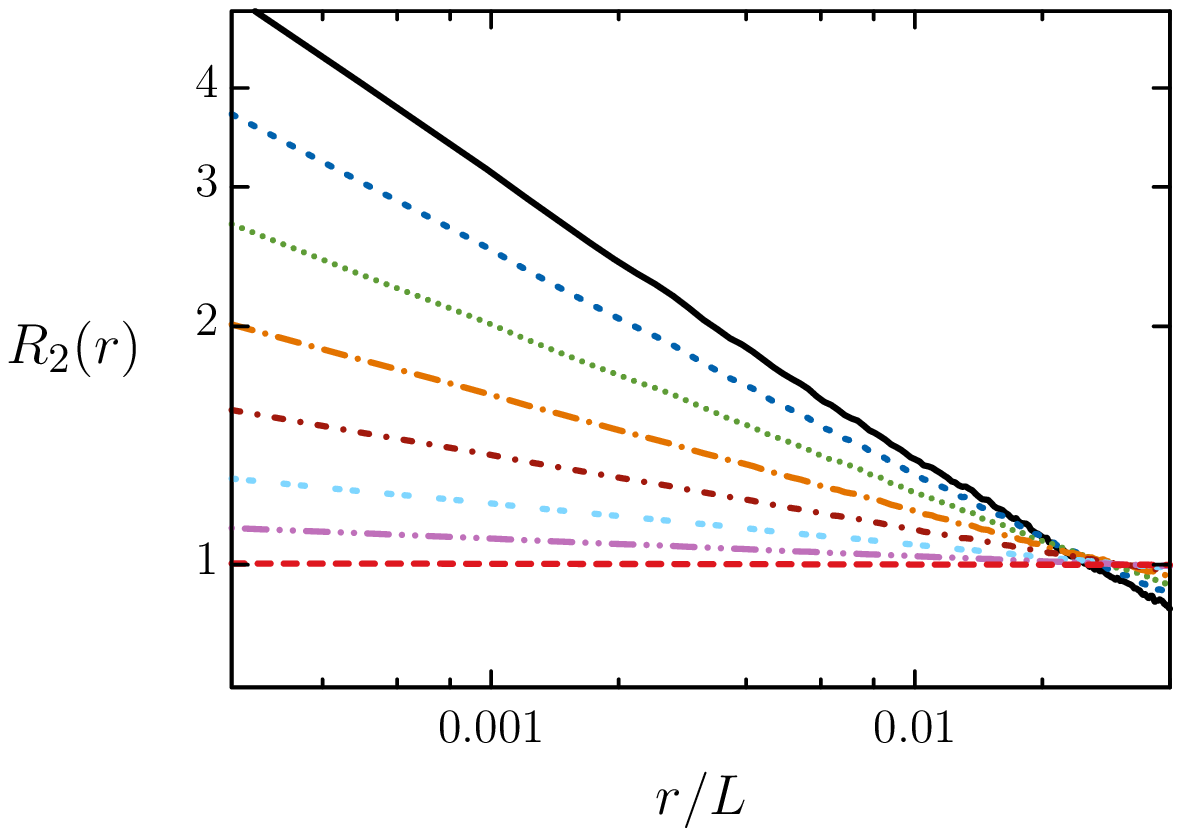}\\[10pt]
\includegraphics*[width=0.44\textwidth]{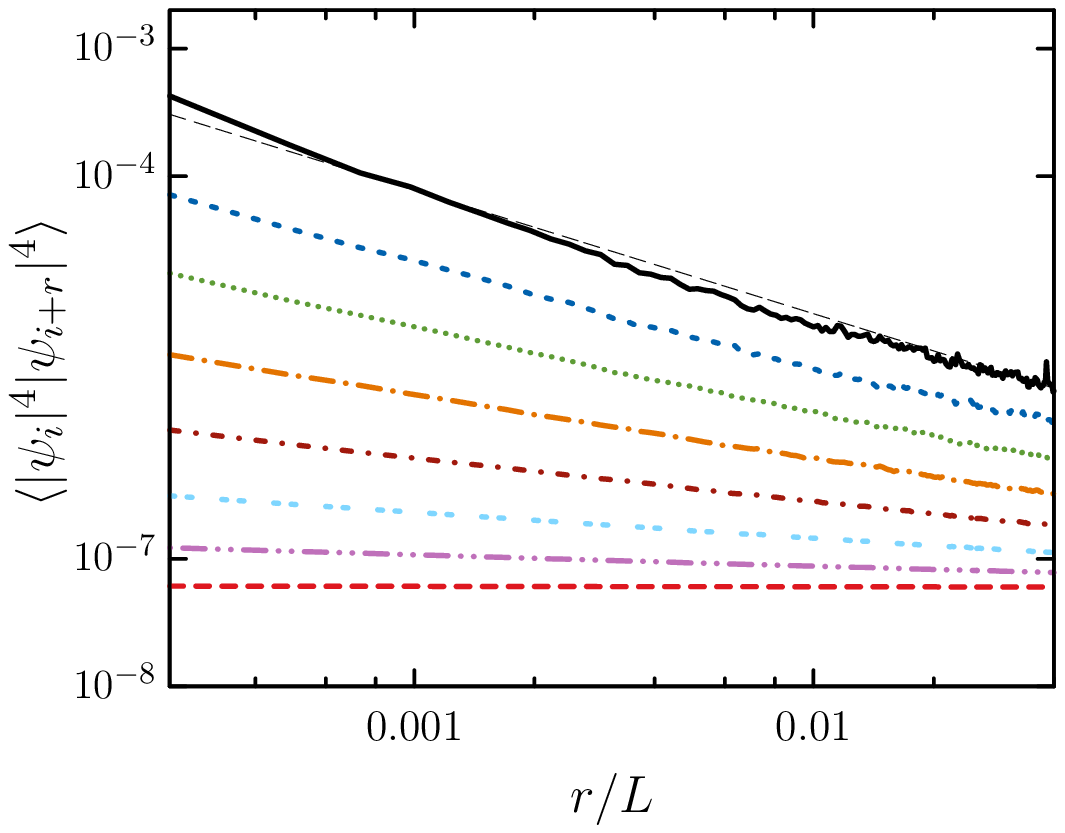}
\caption{(Color online) Correlation functions for the random intermediate map 
of size $N=2^{12}$ in the vicinity of $\gamma = 1/5$. Top: 2-point function (\ref{defR2}), bottom:
higher order correlation function. The black dashed straight line
corresponds to $r^{2D_3-D_2-1}$, as expected from \cite{CatDeu87}.
The curves correspond to $\gamma=1/5+\epsilon/(5N)$ with (from top to bottom) $\epsilon=0$ (black
solid line), $0.127$ (blue dashed line), $0.255$ (green dotted line), $0.382$ (orange
dash-dotted line), $0.509$ (dark-red dash-dotted line), $0.637$ (light-blue double-dashed line),
$0.764$ (purple dash-double dotted line), $0.955$ (red dashed line). 
\label{r2gamma}}
\end{figure}

A crucial property of this type of perturbation, as our numerical results show, is that multifractal properties for different values of $\gamma$ do not depend on any characteristic length. This can be shown by investigating the $2-$point correlation function  \eqref{defR2} for this model. Results displayed in Fig.~\ref{r2gamma} (top) show that $R_2$ behaves as a power law as in Eq.~(\ref{R2_D2}) over a broad range of scales and yields a well-defined multifractal dimension $D_2$ which increases toward the ergodic value $D_2=1$ when the parameter is tuned away from criticality. The fact that multifractal dimensions change smoothly and uniformly at all scales is a footprint of scenario II. To confirm this effect we have also computed higher order correlation functions, which are known to involve other multifractal dimensions \cite{mirlinRMP08}. An example is shown in Fig.~\ref{r2gamma} (bottom), showing that indeed other multifractal dimensions follow scenario II.  

\subsubsection{Outline of the perturbation theory}
As mentioned above, it is possible to build analytically a perturbation theory to describe the vicinity of rational values of $\gamma$. This is done by using a related mathematical model, the Ruijsenaars-Schneider model \cite{ruijsc}. As shown in \cite{BogGir}, this model also displays multifractal eigenvectors. It was already used to describe a quantum version of the intermediate map with unbounded phase space \cite{MGGG12}. 

In this section we develop a perturbation theory for the intermediate map when the parameter $\gamma$ depends on the matrix size. Namely, we consider the intermediate map at parameter value of the form $\gamma=1/b+a/N$, that is, when the parameter of the map gets close to a rational at a speed which depends on $N$. 
The Ruijsenaars-Schneider (RS) ensemble is defined as the ensemble of $N\times N$ unitary matrices of the form
\begin{equation}
U_{mn}=\frac{\mathrm{e}^{\mathrm{i}\Phi_m}}{N}\frac{1-\mathrm{e}^{2\pi\mathrm{ i} g }}{1-\mathrm{e}^{2\pi\mathrm{ i}(m-n+g)/N}},
\label{defruijsenaars}
\end{equation}
with $\Phi_m$ independent random phases uniformly distributed between $0$ and $2\pi$, and $g$ some fixed parameter \cite{BogGir}. 

The perturbation expansion for multifractal dimensions of the Ruijsenaars model was obtained in \cite{BogGir} in the weak multifractality limit where $g$ is close to a nonzero integer. The intermediate map at parameter value $\gamma=1/b+a/N$ coincides with the Ruijsenaars map with parameter $g=N\gamma=N/b+a$. Let $r$ be the remainder of $N$ modulo $b$. The weak multifractality limit for  \eqref{defruijsenaars} is obtained when $N/b+a$ is close to a nonzero integer, that is, when $a=\kappa-r/b+\epsilon$, with $\kappa$ an integer and $\epsilon$ a small real number. For such a value of $a$, the intermediate map corresponds to the map \eqref{defruijsenaars} with parameter $g=\tilde{\kappa}+\epsilon$, where $\tilde{\kappa}=(N-r)/b+\kappa$ is an integer. Thus we expect multifractal dimensions for the intermediate map to be given by a perturbation expansion in $\epsilon$ similar as in \cite{BogGir}, but in the vicinity of an integer $\tilde{\kappa}$ which depends on $N$. For simplicity, we will consider the case where gcd$(\tilde{\kappa},N)=1$. We can then define $\tilde{\kappa}^{-1}$ as the inverse of $\tilde{\kappa}$ modulo $N$.

Multifractal dimensions can be obtained from the asymptotic behavior of 
\begin{equation}
\label{momentmethod}
\sum_n|\Psi_n(\alpha)|^{2q}\sim N^{-D_q(q-1)}
\end{equation}
at large $N$. Here $\Psi_n(\alpha)$ is the nth component of the $\alpha$th eigenvector of the system \eqref{defruijsenaars} of size $N$. In the unperturbed case where states are extended, the multifractal dimensions are $D_q^{(0)}=1$ for all $q$. The perturbative approach allows to express eigenvectors of \eqref{defruijsenaars} at first order in $\epsilon$, and thus the moments of the wavefunction. At $\epsilon>0$, fractal dimensions are given by $D_q=1-\frac{q}{2} d_q$, with $d_q$ some small number (the factor $\frac12$ is put here for convenience). From \eqref{momentmethod} one then  obtains
\begin{eqnarray}
\label{momentsdq}
\sum_n|\Psi_n(\alpha)|^{2q}&\sim& N^{-(1-q d_q/2)(q-1)}\nonumber\\
&\simeq &\frac{1}{N^{q-1}}\left(1+\frac{q(q-1)}{2}d_q \ln N\right).
\end{eqnarray}
The first-order correction to the multifractal dimension is thus given by the logarithmic behavior of the perturbative
correction of the moments. We first find a closed expression for the first-order correction of the wavefunction
moments averaged over the whole spectrum and over disorder configurations (Eqs.~\eqref{psi2qtemp2} and \eqref{big3}), and then extract the
dominant logarithmic contribution in the limit $N\to\infty$ (Eq.~\eqref{sommerandom}), which gives us the correction $d_q$ sought for.\\
Eventually, the calculations detailed in the next paragraph provide an analytical confirmation that a change of $\gamma$ in the intermediate map leads to a multifractality breakdown following scenario II.

\subsubsection{Perturbation expansion}

Let us consider a perturbation expansion of \eqref{defruijsenaars} around $\tilde{\kappa}$, setting $g=\tilde{\kappa}+\epsilon$. Let $M_{mn}=U_{mn}e^{-i\pi \epsilon (1-1/N)}$: this rescales $U_{mn}$ by a trivial factor, and
\begin{eqnarray}
\label{rescale}
M_{mn}&=&\delta_{m-n+\tilde{\kappa}}\frac{e^{i\Phi_m}}{N}\frac{\sin\pi \epsilon}{\sin(\pi \epsilon/N)}\nonumber\\
&+&(1-\delta_{m-n+\tilde{\kappa}})\frac{(1-e^{2\pi i \epsilon })e^{-i\pi \epsilon (1-1/N)}}{1-e^{2\pi i(m-n+\tilde{\kappa}+\epsilon)/N}}
\end{eqnarray}
is then such that both terms have a definite limit when $\epsilon\to 0$. First-order expansion of $M_{mn}$ reads
\begin{equation}
\label{M1RS}
M_{mn}\simeq e^{i\Phi_m}\delta_{m-n+\tilde{\kappa}}-\frac{2i\pi\epsilon}{N}e^{i\Phi_m}\frac{1-\delta_{m-n+\tilde{\kappa}}}{1-e^{2\pi i(m-n+\tilde{\kappa})/N}}.
\end{equation}
We denote eigenfunctions and eigenvalues of $M_{mn}$ respectively by $\Psi_{n}(\alpha)$ and $\lambda_{\alpha}$. Unperturbed eigenstates, that is, eigenvectors of $e^{i\Phi_m}\delta_{m-n+\tilde{\kappa}}$, are given by
\begin{eqnarray}
\Psi^{(0)}_n(\alpha)&=&\frac{1}{\sqrt{N}}e^{i S_{\tilde{\kappa}^{-1}n}(\alpha)},\nonumber\\
 S_n(\alpha)&=&\frac{2\pi}{N}n\alpha+n\tilde{\Phi}-\sum_{j=0}^{n-1}\Phi_{\tilde{\kappa}j}
\label{psi0}
\end{eqnarray}
with eigenvalues
\begin{equation}
\lambda_{\alpha}^{(0)}=e^{i \tilde{\Phi}+\frac{2i\pi}{N}\alpha},
\end{equation}
where $\tilde{\Phi}=\frac1N\sum_{j=0}^{N-1}\Phi_j$. Standard first-order perturbation expansion of eigenvectors gives
\begin{equation}
\Psi_n(\alpha)=\Psi^{(0)}_n(\alpha)+\sum_{\beta}C_{\alpha \beta}\Psi^{(0)}_n(\beta),
\label{Psi_n_alpha}
\end{equation}
with
\begin{equation}
\label{defC}
C_{\alpha \beta}=\frac{\sum_{mn}\Psi^{(0)*}_m(\beta)M^{(1)}_{mn}\Psi^{(0)}_n(\alpha)}{\lambda_{\alpha}^{(0)}-\lambda_{\beta}^{(0)}}
\end{equation}
and $M^{(1)}_{mn}$ the order-$\epsilon$ (off-diagonal) term in \eqref{M1RS}. Replacing $\Psi^{(0)}_n$ by its explicit value \eqref{psi0} in Eq.~\eqref{Psi_n_alpha} we get 
\begin{equation}
|\Psi_n(\alpha)|^2=\frac{1}{N}\left(1+Q_n(\alpha)+Q_n^{*}(\alpha)+Q_n(\alpha)Q_n^{*}(\alpha)\right)
\label{psi2bis}
\end{equation}
with 
\begin{equation}
\label{defqn}
Q_n(\alpha)=\sum_{\beta}e^{2i\pi\beta n/N}C_{\alpha, \beta+\alpha}.
\end{equation}
Taking \eqref{psi2bis} to the power $q$ and summing over $n$ we get, up to order $\epsilon^2$,
\begin{equation}
\label{psi2qtemp2}
\sum_n|\Psi_n(\alpha)|^{2q}=\frac{1}{N^{q-1}}+\frac{q(q-1)}{2N^q}\sum_n(Q_n(\alpha)+Q_n^{*}(\alpha))^2 ,
\end{equation}
where terms linear in $q$ sum up to zero because of the normalization of $\Psi$. Identifying \eqref{psi2qtemp2} and \eqref{momentsdq}, we get 
\begin{equation}
\label{sommeruij}
\frac{1}{N}\sum_n(Q_n(\alpha)+Q_n^{*}(\alpha))^2\simeq d_q \ln N,
\end{equation}
so that the correction to the unperturbed multifractal dimension is indeed given by the logarithmic asymptotic behavior of the sum in \eqref{sommeruij}.

\subsubsection{Model with random phases}
From the definition \eqref{defqn} of $Q_n$ we have
\begin{equation}
\frac{1}{N}\sum_nQ_n(\alpha)^2=\sum_{\beta}C_{\alpha, \alpha+\beta}C_{\alpha, \alpha-\beta},
\label{sommeQQ}
\end{equation}
and 
\begin{equation}
\frac{1}{N}\sum_nQ_n(\alpha)Q^{*}_n(\alpha)=\sum_{\beta}C_{\alpha, \alpha+\beta}C^{*}_{\alpha, \alpha+\beta}.
\label{sommeQQstar}
\end{equation}
We are interested in quantities averaged over all eigenvectors and random phases: we thus have to perform a sum over $\alpha$ and an integral over the $\Phi_j$. From Eqs.~\eqref{M1RS}--\eqref{defC}, the explicit expression of $C_{\alpha, \beta+\alpha}$ is
\begin{eqnarray}
\label{caba}
&C_{\alpha, \beta+\alpha}=\frac{i}{2N}\sum_{rs}\frac{t_{(s-r+1)\tilde{\kappa}}}{\sin\frac{\pi\beta}{N}}
\exp\left[i(r-s-1)(\frac{2\pi\alpha}{N}+\tilde{\Phi})\right]\nonumber\\
&\hspace{-.5cm}\times\exp\left[-\frac{2i\pi}{N}(s+\frac12)\beta-i\sum_{j=0}^{r-1}\Phi_{\tilde{\kappa}j}+i\sum_{j=0}^{s}\Phi_{\tilde{\kappa}j}\right]
\end{eqnarray}
(we have changed the summation from $m,n$ to $r,s$ with $r= \tilde{\kappa}^{-1}n$ and $s=\tilde{\kappa}^{-1}m$), and
\begin{equation}
\label{deft}
t_x=\frac{\pi\epsilon}{N}\frac{e^{-i\pi x/N}}{\sin\pi x/N}\quad\textrm{if $x\neq 0$,}\qquad 
\textrm{$0$ otherwise.}
\end{equation}
Similarly $C_{\alpha, \alpha-\beta}$ can be expressed by a sum over indices $r'$ and $s'$. The quantity $C_{\alpha, \beta+\alpha}C_{\alpha, \alpha-\beta}$ depends on $\alpha$ through a factor $\exp[i(r-s-1+r'-s'-1)\frac{2\pi\alpha}{N}]$. Upon averaging over $\alpha$, one thus gets a coefficient $\delta_{r-s+r'-s'-2}$ which kills the terms $\tilde{\Phi}$. The averaging of $C_{\alpha, \alpha+\beta}C_{\alpha, \alpha-\beta}$ over random angles then contains a coefficient
\begin{equation}
\label{sommeangles}
\Big\langle\exp \left[-i\sum_{j=0}^{r-1}\Phi_{\tilde{\kappa}j}+i\sum_{j=0}^{s}\Phi_{\tilde{\kappa}j}-i\sum_{j=0}^{r'-1}\Phi_{\tilde{\kappa}j}+i\sum_{j=0}^{s'}\Phi_{\tilde{\kappa}j}\right]\Big\rangle.
\end{equation}
Since $r-s+r'-s'-2=0$, and $t_0=0$ (so that contributions with $r=s+1$ or $r'=s'+1$ vanish), the average \eqref{sommeangles} can only be nonzero when $s=r'-1$ and $s'=r-1$. This yields
\begin{equation}
\label{CCrandom}
\Big\langle C_{\alpha, \alpha+\beta}C_{\alpha, \alpha-\beta}\Big\rangle_{\alpha, \Phi}\hspace{-.2cm}=
\frac{-1}{4N^2}\sum_{rs}\frac{|t_{(s-r+1)\tilde{\kappa}}|^2}{\sin^2\frac{\pi\beta}{N}}
e^{-\frac{2i\pi}{N}(s-r+1)\beta}. 
\end{equation}
Changing variables $s-r+1=x$ and summing over $\beta$, we get from \eqref{sommeQQ} and \eqref{CCrandom}
\begin{equation}
\label{sumqq}
\Big\langle\frac{1}{N}\sum_nQ_n(\alpha)^2\Big\rangle_{\alpha, \Phi}=
-\frac{\pi^2\epsilon^2}{4N^3}\sum_{x,\beta}\frac{1}{\sin^2\frac{\pi \tilde{\kappa} x}{N}}\frac{e^{-\frac{2i\pi}{N}x\beta}}{\sin^2\frac{\pi\beta}{N}}.
\end{equation}
In a similar way, averaging $C_{\alpha, \alpha+\beta}C^{*}_{\alpha, \alpha+\beta}$ over $\alpha$ yields a coefficient $\delta_{r-s-r'+s'}$, and the average over $\Phi_j$ then yields the condition $r=r'$ and $s=s'$. An expression equivalent to \eqref{sumqq} can be found, which, summed together with Eq.~\eqref{sumqq}, yields
\begin{equation}
\Big\langle\frac{1}{N}\sum_n(Q_n(\alpha)+Q_n^{*}(\alpha))^2\Big\rangle_{\alpha, \Phi}=\frac{\pi^2\epsilon^2}{N^3}\sum_{x}\frac{x(N-x)}{\sin^2(\pi \tilde{\kappa} x/N)},
\label{big3}
\end{equation}
where we have performed the sum over $\beta$ by using the identity
\begin{equation}
\sum_{\beta=1}^{N-1}\frac{\sin^2(\pi\beta x/N)}{\sin^2(\pi\beta/N)}=x(N-x).
\end{equation}
From \eqref{sommeruij} we know that the lowest-order contribution to the multifractal exponent is given by the logarithmic behavior of the sum \eqref{big3} for large $N$. Recall that $\tilde{\kappa}=(N-r)/b+\kappa$ with $\kappa$ fixed. The sum in \eqref{big3} can be split into $b$ subsums, as
\begin{equation}
\label{splitsum}
\frac{\pi^2\epsilon^2}{N^3}\sum_{c=0}^{b-1}\sum_{x=0}^{\lfloor N/b\rfloor-1}\frac{(bx+c)(N-bx-c)}{\sin^2\pi\left(\frac{c}{b}+\frac{(\kappa b-r)(bx+c)}{Nb}\right)}
\end{equation}
(we obviously omit the case $c=x=0$ in the above sum). The logarithmic contribution \eqref{sommeruij} originates from regions where the divergence of the $\sin^2$ in the denominator is compensated by a linearly vanishing numerator. This corresponds to the two regions $b x+c\simeq 0$ and $b x+c\simeq N$ (in all other cases, either the summand is $\sim 1/x^2$ and converges, or it gives non-logarithmic divergences which should be compensated by higher-order terms in the perturbation expansion if we assume that the behavior \eqref{sommeruij} holds). The first region comes from the sum with $c=0$ in \eqref{splitsum}, the second one from the sum with $c=r$.

The sum for $c=0$ in \eqref{splitsum} can be rewritten as the sum of two terms, namely
\begin{equation}
\label{sommec0bis}
\frac{\pi^2\epsilon^2}{N^3}\sum_{x=1}^{\lfloor N/b\rfloor-1}\left(\frac{bx(N-bx)}{\sin^2\pi\left(\frac{(\kappa b-r)}{N}x\right)}-\frac{b N^3}{(\kappa b-r)^2\pi^2x}\right),
\end{equation}
 which is a Riemann sum converging to an integral with a finite value, and 
\begin{equation}
\frac{b \epsilon^2}{(\kappa b-r)^2}\sum_{x=1}^{\lfloor N/b\rfloor-1}\frac{1}{x}\simeq
\frac{b \epsilon^2}{(\kappa b-r)^2}\ln N
\end{equation}
which is responsible for the logarithmic divergence. After a change of variables $x\rightarrow (N-r)/b-x$, the sum for $c=r$ in \eqref{splitsum} can be shown to yield exactly the same contribution as $c=0$. Summing both contributions, one gets 
\begin{equation}
\label{sommerandom}
\Big\langle\frac{1}{N}\sum_n(Q_n(\alpha)+Q_n^{*}(\alpha))^2\Big\rangle_{\alpha, \Phi}\simeq
\frac{2b \epsilon^2}{(\kappa b-r)^2}\ln N,
\end{equation}
which, by identification with \eqref{sommeruij}, gives $d_q$ and thus $D_q$. Since $\gamma=1/b+(\kappa-r/b+\epsilon)/N$, one finally gets
\begin{equation}
\label{dqfinalrandom}
D_q\simeq 1- \frac{q}{b}\left(1-N\frac{\gamma b-1}{\kappa b-r}\right)^2.
\end{equation}
The first maxima of $D_q$ around $\gamma=1/b$ correspond to $\kappa=0, \pm 1, \pm 2\ldots$ The dependence in $N$ in Eq.~\eqref{dqfinalrandom} indicates that this fractal dimension $D_q$ gives the behavior of wavefunctions of size $N$ when the parameter $\gamma$ is considered at a scale $1/N$ around the rational $1/b$.

The theory can be compared with the numerical results; Fig.~\ref{Dq_vsg} shows that indeed it describes correctly the vicinity of rational points for finite $N$ for positive or negative $q$.

\subsubsection{The case of deterministic phases}

\begin{figure}[t]
\includegraphics*[width=0.48\textwidth]{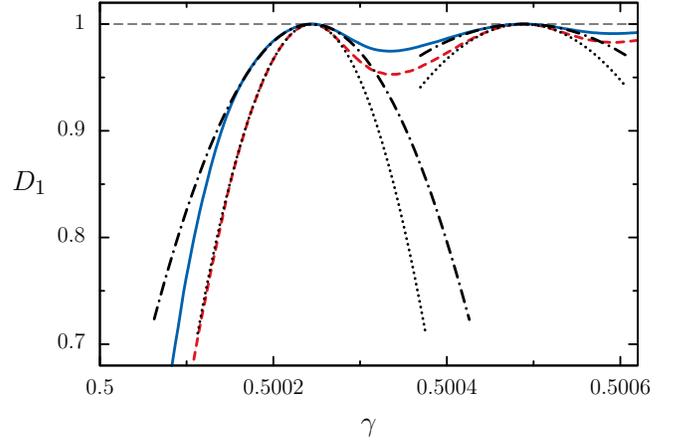}
\caption{(Color online) Multifractal dimensions $D_{1}(\gamma)$ for the intermediate map  with random phases (blue
solid line) and deterministic phases (red dashed line) in the vicinity of $\gamma = 1/2$ for
$N=2^{12}$. Black dash-dotted parabolas correspond to  the theoretical expression Eq.~\eqref{dqfinalrandom} for $\kappa=1,2$ while black dotted parabolas
correspond to Eq.~\eqref{dqfinalrandom} with an additional prefactor $2$ in front of $q$. 
\label{gammap2}}
\end{figure}
\begin{figure}[t]
\includegraphics*[width=0.45\textwidth]{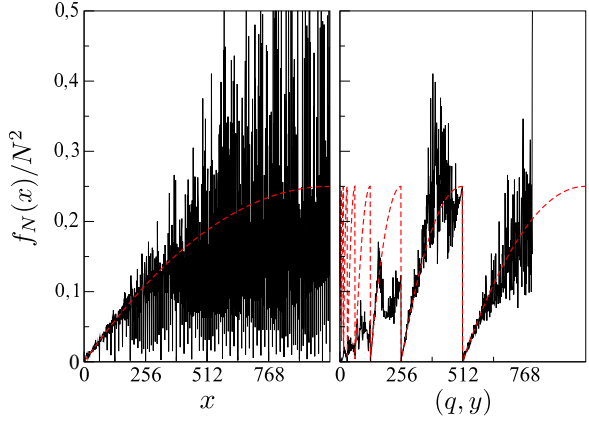}
\caption{(Color online) Left: Function $f_N(x)$ defined in \eqref{deffn} (see text) as a function of $x$ for $N=2^{11}$ (black), and function $x(N-x)$ (dashed/red). Right: Same plot with a different ordering of the abscissas. Points with abscissa $x=2^{n-q}(2y+1)$ with $q=1,2,\ldots,n$ and $y=0,1,\ldots, 2^{q-2}$ are ordered by increasing $q$ and at fixed $q$ by increasing $y$. The last family (from 512 to 1024) corresponds to $q=n$, that is, odd $x$.
\label{rand_vs_p2}}
\end{figure}

It is instructive to also analyze the fractal dimensions in the case where random phases are replaced by the phases $\Phi_k=2\pi k^2/N$ of the deterministic intermediate model. Numerically, this model yields results which are close but slightly different from those of the model with random phases. While the difference is tiny for most values of the denominator $b$ of the parameter, the most significant discrepancy occurs in the case $\gamma=1/2$. For instance, as illustrated in Fig.~\ref{gammap2}, when $\gamma$ is equal to $1/2$ and $N$ is a power of 2, the value of $1-D_1$ for deterministic phases is twice the value for random phases given by Eq.~\eqref{dqfinalrandom}. The method used above can in fact be adapted to explain these discrepancies. In this subsection we will show this for $\gamma=1/2$ and $N=2^n$. Other cases can be treated by a similar approach.


To obtain the result for the deterministic model, the main change in the perturbative calculation of $D_q$ comes from Eq.~\eqref{sommeangles}, where the average over random phases is replaced by a constant term $\xi_{rs}\xi_{r's'}$, with
\begin{equation}
\label{sommeanglesP2}
\xi_{rs}=\exp\left[\frac{2i\pi\tilde{\kappa}^2}{N}\left(-\sum_{k=0}^{r-1}k^2+\sum_{k=0}^{s}k^2\right)\right].
\end{equation}
Performing the same steps as previously, one can show that
\begin{equation}
\label{QQp3}
\Big\langle\frac{1}{N}\sum_n(Q_n(\alpha)+Q_n^{*}(\alpha))^2\Big\rangle_{\alpha}
=\frac{\pi^2\epsilon^2}{N^3}\sum_x\frac{f_N(x)}{\sin^2(\pi \tilde{\kappa}x/N)},
\end{equation}
where the numerator $x(N-x)$ of Eq.~\eqref{big3} is now replaced by a function
\begin{eqnarray}
\label{deffn}
f_N(x)&=&\frac{1}{2N}\sum_{rr'ss'}\sigma(s'-s)\delta_{r-s-1-x}\\
&\times&\textrm{Re}\left[
\xi_{r,s}(-\xi_{r',s'}\delta_{r'-s'-1+x}+\xi_{r',s'}^*\delta_{r'-s'-1-x})\right],\nonumber
\end{eqnarray}
with $\sigma$ an $N$-periodic function defined for $0\leq v\leq N-1$ by
\begin{equation}
\sigma(v)\equiv\sum_{\beta}\frac{e^{\frac{2i\pi}{N}v\beta }}{\sin^2(\pi\beta/N)}=\frac{N^2-1}{3}-2v(N-v).
\end{equation}
As before, the logarithmic behavior of \eqref{QQp3} is expected to come from places where $\sin^2(\pi \tilde{\kappa} x/N)$ becomes close to zero while the numerator approaches zero linearly. Previously, when the numerator was given by $x(N-x)$, this only occurred for $x\simeq 0$ and $x\simeq N$. Here the function $f_N(x)$ still has a linear behavior in the vicinity of $x\simeq 0$ but it is much more oscillating for larger $x$. For illustration, we plot an example of $f_N(x)$ in Fig.~\ref{rand_vs_p2} (left panel) for $N=2^n$; we concentrate on indices $x\in[0,N/2]$ since by symmetry the other half yields the same contribution. The only linear contribution would seem to come from $x\simeq 0$. 
However, let us consider the case $N=2^n$ and $\gamma=1/2$. If we rearrange the labels $x$ by setting  $x=2^{n-q}(2y+1)$ with $q=1,2,\ldots,n$ and $y=0,1,\ldots, 2^{q-2}$, we obtain the right panel in Fig.~\ref{rand_vs_p2}. We see that the linear behavior of $f_N(x)$ does not occur only for $x\simeq 0$, but also for all families $x=2^{n-q}(2y+1)$  when $y\simeq 0$. Namely, for all $q$ and $y$ small we observe numerically that
\begin{equation}
\label{fnxy}
f_N(x)\simeq x(N-x)=2^{2(n-q)}(2y+1)(2^q-2y-1).
\end{equation}
For $\tilde{\kappa}=N/2+1$ and even $x$, i.e.~families $x=2^{n-q}(2y+1)$ with $1\leq q\leq n-1$, we have
\begin{equation}
\label{sinxy}
\sin^2\frac{\pi\tilde{\kappa} x}{N}=\sin^2\left[\pi\left(\frac{1}{2}+\frac{1}{N}\right)x\right]
=\sin^2\frac{\pi(2y+1)}{2^q},
\end{equation}
so that \eqref{fnxy} will partly compensate \eqref{sinxy}, and thus for small $y$ each $q$-family with $q\neq n$ will give a logarithmic contribution to \eqref{QQp3}. As before, odd $x$ do not contribute, since for $x=2y+1$ (the family $q=n$) we get
\begin{equation}
\label{cosxy}
\sin^2\frac{\pi\tilde{\kappa} x}{N}=\cos^2\frac{\pi x}{N}=\cos^2\frac{\pi(2y+1)}{2^n},
\end{equation}
so that vanishing of \eqref{cosxy} does not correspond to a vanishing of \eqref{fnxy}. Using \eqref{fnxy} and \eqref{sinxy}, the contribution to \eqref{QQp3} of a  $q$-family $x=2^{n-q}(2y+1)$ with $1\leq q\leq n-1$ for small $y$ reads  
\begin{equation}
\label{sommeq}
\frac{\pi^2\epsilon^2}{N^3}\sum_y\frac{2^{2(n-q)}(2y+1)(2^q-2y-1)}{\sin^2\frac{\pi(2y+1)}{2^q}}
\simeq \frac{\epsilon^2}{N}2^q\sum_y\frac{1}{2y+1}.
\end{equation}
Then for large $q$
\begin{equation}
\label{sommeq2}
\sum_{y=0}^{2^{q-2}}\frac{1}{2y+1}=\sum_{y=1}^{2^{q-1}+1}\frac{1}{y}-\sum_{y=1}^{2^{q-2}}\frac{1}{2y}
\sim\frac{1}{2}\ln 2^q,
\end{equation}
and the sum over all contributions \eqref{sommeq} gives 
\begin{equation}
\label{oscil}
\frac{\epsilon^2}{2N}\sum_{q=1}^{n-1}2^q\ln 2^q\sim \frac{\epsilon^2}{2}\ln N.
\end{equation}
So far we considered only the region  $x<N/2$. The region $x\in[N/2,N]$ contributes in the same way, so that the total contribution from the oscillating part of $f_N(x)$ is twice the result \eqref{oscil}, that is, $\epsilon^2\ln N$. To this contribution one must add the contribution from the linear part of $f_N(x)$, corresponding to $x\simeq 0$ (and $x\simeq N$), see Fig.~\ref{rand_vs_p2} left; the calculation is the same as for random phases and thus yields the contribution given by Eq.~\eqref{sommerandom}. For $b=2$ and $\kappa=1$ this term is equal to $\epsilon^2\ln N$. The total of all contributions for deterministic phases is thus $2\epsilon^2\ln N$, i.e.~twice the total for random phases. Figure \ref{gammap2} illustrates this factor 2 between random and deterministic phases.



\section{Change of measurement basis}
\label{basis_change}

An intriguing characteristics of multifractal properties is their dependence on the basis 
choice. Indeed, it is known for the intermediate map \cite{Giraud} that multifractal properties for 
rational $\gamma$, which are visible in the momentum basis, disappear in the position basis. It 
is all the more surprising that a recent conjecture \cite{BogGir} proposes to link the spectral 
statistics (independent of the basis) to the multifractal spectrum (a priori basis dependent). 
Apart from its fundamental interest, this question is also important for experimental 
implementations. Indeed, it is not always evident to choose the measurement basis at will in an 
experiment, and it is thus interesting to assess how multifractality is modified when different 
observables are used. The main idea of this section is thus to identify how the multifractality 
spectrum varies when the basis is changed. 

The results of our analysis have shown that in this case the multifractality breakdown follows the broad picture of scenario II where the multifractality at small scales is uniformly destroyed. However, we have found that this broad picture admits several variants depending on the presence or absence of a characteristic length in the model itself or in the perturbation. In the absence of a perturbation, both the PRBM and the Anderson models have no characteristic length, while the intermediate map exhibits such a length (see Section II). When no characteristic length is present in the unperturbed model,  like for the Anderson model, we were able to construct two kinds of perturbations, which themselves may or may not exhibit an intrinsic characteristic length.

In the following, we first discuss in Subsection V.A the PRBM model and the Anderson model in the case where the perturbation does not introduce a characteristic length, showing that in this case the results correspond to scenario II. However, in Subsection V.B we consider the intermediate map and the Anderson model when the perturbation has a characteristic length. In both cases, we show through a two-parameter scaling analysis the presence of a perturbation-dependent characteristic length, below which the multifractality is uniformly destroyed (following scenario II). However, as we shall see, the behavior is different above the characteristic scale.

\subsection{Absence of a characteristic length}

\subsubsection{PRBM model}

We first consider the PRBM model defined by (\ref{PRBMvarcoeff}). We construct a generic change of basis through a smooth deformation of the identity. The unitary matrix defining the basis change is chosen to be
\begin{equation}
  \label{deformId}
  U(\epsilon)=e^{\ic \epsilon M}
\end{equation}
where $\epsilon$ is the deformation parameter and $M$ an element of the Gaussian Orthogonal Ensemble (GOE) of random matrices. A matrix $H$ of the PRBM in the new basis becomes $H'$ given by:
\begin{equation}
\label{deformId2}
 H'=U(\epsilon)\, H\, U(\epsilon)^{-1}.
\end{equation}
 In order to get generic results, we average over a sample of matrices $M$ from GOE.
\begin{figure}[ht]
    \centering
      \includegraphics*[width=.45\textwidth]{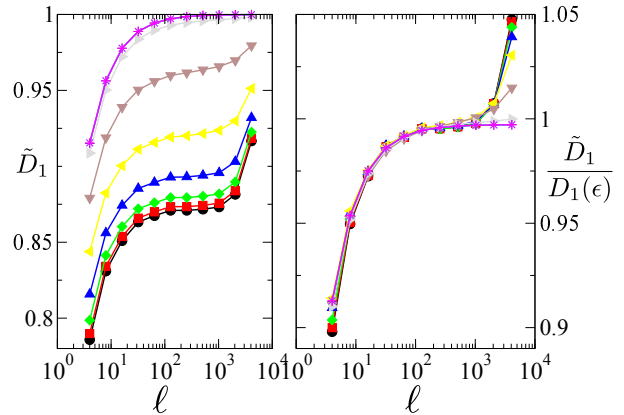}
  \caption{(Color online) Scaling analysis of the local multifractal dimension $\tilde{D}_1$ for the PRBM model (\ref{PRBMvarcoeff}) for $N=8192$ and $b=1$ for different change of basis (\ref{deformId2}). Left: raw data. Right: data after vertical rescaling of $\tilde{D}_1$ by a factor $D_1(\epsilon)$.
  Each curve corresponds to a different value of the parameter $\epsilon$ in (\ref{deformId}). It is chosen to be of the form $\epsilon=10^{0.1 n-3}$; black circles: $n=1$; red squares: $n=3$; green diamonds: $n=5$; blue up triangles: $n=7$; yellow left triangles: $n=9$; brown down triangles: $n=11$; gray right triangles: $n=13$; purple pluses: $n=15$; cyan crosses: $n=17$; magenta stars: $n=19$. $10$ realizations of GOE matrices in (\ref{deformId}) are taken in order to average over $81920$ vectors.
}
  \label{Dqlocal_PRBM}
\end{figure} 
 
\begin{figure}[ht]
    \centering
     \includegraphics*[width=.45\textwidth]{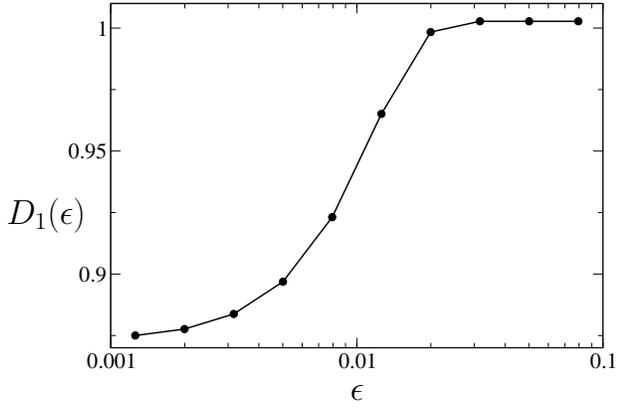}
  \caption{Scaling parameter $D_1(\epsilon)$ extracted from the analysis of Fig.~\ref{Dqlocal_PRBM}. It is normalized to correspond to the multifractal dimension in the PRBM model (\ref{PRBMvarcoeff}) with $N=2^{13}$, $b=1$ in the limit $\epsilon\to 0$.  The perturbation parameter $\epsilon$ quantifies a generic change of basis (\ref{deformId2}).
}
  \label{D1_vs_epsPRBM}
\end{figure}

Fig.~\ref{Dqlocal_PRBM} displays the curves $\tilde{D}_1(\ell)$ of the eigenvectors for different $\epsilon$, showing that an appropriate {\it vertical} rescaling enables to collapse them (as opposed to the previous {\it horizontal} rescaling performed in Section III). The rescaling here just affects the height of the plateau and not the scale at which it appears, clearly indicating that scenario II is followed: the multifractality disappears uniformly when the perturbation is increased. As Fig.~\ref{Dqlocal_PRBMbis} shows at $\epsilon=0$, multifractal dimensions are given by the plateau appearing at intermediate scales, and therefore the rescaling should be made on these ranges of scales.  The scaling parameter $D_1(\epsilon)$ (corresponding to the mean value of the plateau) is displayed in Fig.~\ref{D1_vs_epsPRBM}, showing that the multifractal dimension smoothly goes to the ergodic value for large $\epsilon$. It was checked (data not shown) that the same scenario also applies for different values of $q$. 

\subsubsection{Anderson model}

\begin{figure}
\includegraphics*[width=.45\textwidth]{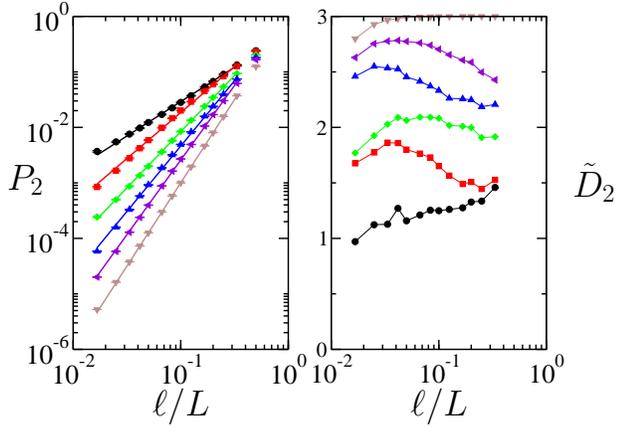}
\caption{(Color online) Left: Moments $P_2$ of the Anderson model with basis change \eqref{eq:UKR} for different values of $t/\tau_\text{Th}$, the Thouless time associated with the quasiperiodic kicked rotor. $W=16.53\approx W_c$, $L=120$ and  $t/\tau_\text{Th}=0$ (black circles), $8.9 \ 10^{-11}$ (red squares), $1.1 \ 10^{-9}$ (green diamonds),  $2.8 \ 10^{-7}$ (blue triangles up),  $8.7 \ 10^{-6}$ (purple triangles tilted),  $2.3$ (brown triangles down). Right: Local multifractal dimensions $\tilde{D}_2$ as a function of $\ell/L$ for different values of $t/\tau_\text{Th}$, same parameters and color code as Left. In both figures each color corresponds to a different couple of $t$ and $K$; 
unperturbed Anderson model (black circles); $t=10$, $K=21$ (red squares); $t=10$, $K=69.5$ (green diamonds); $t=21$, $K=760.9$ (blue triangles up); $t=59$, $K=2517.7$ (purple triangles tilted); $t=100$, $K=998784$ (brown triangles down). $\tau_\text{Th}\equiv N^2/D$ was determined by a numerical determination of the diffusion constant $D$ of the 1D quasiperiodic kicked rotor.}
\label{fig:IPRvslboxovL}
\end{figure}

Here, we investigate a change of basis for the Anderson model. We consider the effects on the multifractal spectrum of a rotation $U$ of the critical states of the Anderson model. A perturbation of the form \eqref{deformId} would require to calculate the exponential of a full matrix $M$ of size $L^3$ for a 3D system up to $L=120$. We will use instead the unitary evolution operator associated with the so-called quasiperiodic kicked rotor \cite{shepelyansky1983some}:
\begin{equation}\label{eq:HKR}
 H_{\mathrm{KR}} = \frac{p^2}{2}+ \mathcal K(t) \cos x \sum_n \delta(t-n) \; ,  
\end{equation}
with $\mathcal K(t)= K [1 + \eta \cos(\omega_2 t) \cos(\omega_3 t)]$ a quasiperiodic kicking amplitude with two frequencies $\omega_2 =2 \pi \sqrt{5}$ and $\omega_3 = 2 \pi \sqrt{13}$ incommensurate with $2 \pi$, $K$ the stochasticity parameter, $t$ the time, $p$ the momentum conjugate to the position $x$. This 1D system is a variant of the famous periodic kicked rotor \cite{casati1979stochastic, grempel1984quantum} (obtained with $\eta=0$), a paradigm of quantum chaos known to exhibit the phenomenon of dynamical localization, i.e. Anderson localization in momentum space. Due to the additional incommensurate frequencies, the quasiperiodic kicked rotor performs an Anderson transition between a localized phase at small $K<K_c$ and a diffusive metallic phase at large $K>K_c$ \cite{casati1989anderson, chabe2008experimental, lemarie2009observation, lemarie2010critical, lopez2013phase}. The evolution operator associated with \eqref{eq:HKR} over a unit step of time writes:
\begin{equation}\label{eq:UKR}
 U= e^{-i \frac{p^2}{2 \hbar}} e^{-i \frac{\mathcal K(t) \cos x}{\hbar}} \;,
\end{equation}
where the value of the effective Planck constant is taken as $\hbar=2.89$ \cite{lemarie2009observation}. We have used this evolution operator \eqref{eq:UKR} in the diffusive metallic phase $K \gg K_c\approx 4.7$ for $\eta=0.8$ to rotate the eigenstates $\vert \Psi(\alpha) \rangle$ of the 3D Anderson model \eqref{eq:HAnd}. More precisely, we have considered $\langle i \vert \Psi(\alpha) \rangle=\Psi_i(\alpha)$ as a vector of size $N=L^3$ in the $p$-space of the quasiperiodic kicked rotor and have made it evolve using the evolution operator $U$ over a time $t$.

\begin{figure}
\includegraphics*[width=.45\textwidth]{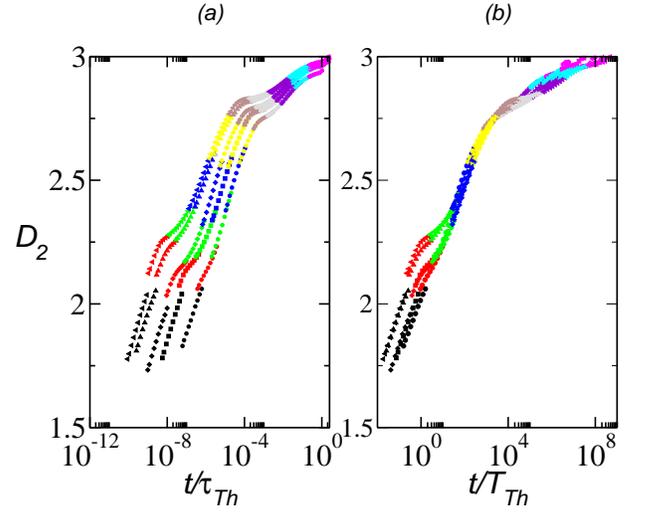}
\caption{
(Color online) Multifractal dimension $D_2$ for the Anderson model with basis change \eqref{eq:UKR} as a function of (a) $t/\tau_\text{Th}$ associated to the 1D quasiperiodic kicked rotor and (b) $t/T_\text{Th}$ associated with the 3D Anderson model, for different sizes $L$ from $L=40$ to $L=120$. The scaling $t/T_\text{Th}$ put the curves for different system sizes on top of each other. However, the ergodic limit $D_q=3$ is reached when $t/\tau_\text{Th} \approx 1$ associated with the 1D dynamics. This arises due to the 1D character of our rotation. 
Each symbol correspond to a different size $L$; from right to left $L=40$ (circles), $60$ (squares), $80$ (diamonds), $100$ (up triangles), $120$ (left triangles). Each color correspond to a different value of $K\equiv \sqrt{2L^6  \hbar^2/\tilde{\tau}}$ with $\tilde{\tau}\equiv\tilde{\tau}_\text{max}  {\tilde{\tau}_\text{min}}^{n/9}/{\tilde{\tau}_\text{max}}^{n/9}$, $n$ an integer varying from $n=0$ to $n=9$ (bottom to top, i.e. black to magenta) and $\tilde{\tau}_\text{max}=L^6/(21^2/2\hbar^2)$, $\tilde{\tau}_\text{min}=50$. For each couple of symbol and color, different points correspond to different times; $t=10$, $12$, $16$, $21$, $27$, $35$, $46$, $59$, $77$ and $100$ from bottom to top. $\tau_\text{Th}\equiv N^2/D$ and $T_\text{Th}=L^2/D$ were determined by a numerical determination of the diffusion constant $D$ of the 1D quasiperiodic kicked rotor.
}
\label{fig:Dqvstovtthouless-L20-120}
\end{figure}

Figure~\ref{fig:IPRvslboxovL} represents the second moment $P_2$ as a function of the box size for different changes of basis, i.e. different values of the diffusion constant $D$ of the quasiperiodic kicked rotor and different evolution times. In the left panel, $P_2$ seems to scale as a power law of the box size $\ell$ over the whole range accessible. If we  study more carefully the local multifractal dimensions $\tilde{D}_2$ (right panel), they show approximate plateaus with small variations but no systematic change which could indicate the presence of a characteristic length.  We also see that the dimension $D_2$ defined as the average of $\tilde{D}_2$ over $\ell$ increases towards the value $D_2=3$ when $t/\tau_\text{Th} \rightarrow 1$, with $\tau_\text{Th}$ the Thouless time associated with the quasiperiodic kicked rotor, $\tau_\text{Th}=N^2/D$, $N=L^3$ and $D$ the diffusion constant of the quasiperiodic kicked rotor. In the case $t/\tau_\text{Th} \approx 1$, the rotated eigenstates are uniformly distributed over the entire sample. 

\begin{figure}
\includegraphics*[width=.45\textwidth]{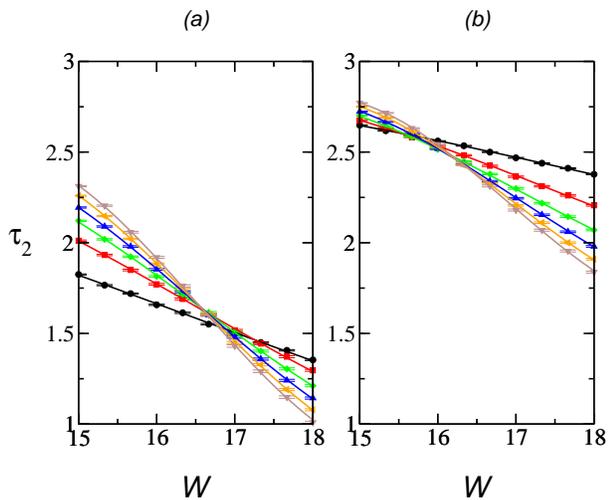}
\caption{(Color online) Anderson transition for the Anderson model with change of basis. Finite-size scaling analysis of $\tau_2=\ln P_2^\text{typ}/\ln \lambda$ for fixed $\lambda=\ell/L=0.1$ as a function of disorder at various system sizes $L \in [20,120]$. The results for the standard Anderson model are represented in (a) while (b) corresponds to the rotated eigenstates with $K=115$ and fixed $t/T_\text{Th}\approx 40$. The error bars are standard deviations. The lines are the fit with a Taylor expansion of the scaling function $\tau_2 = F(L/\xi)+L^{-y} F_{\textrm{irr}}(L/\xi)$, and $\xi\sim\vert W-W_c \vert^{-\nu}$ the scaling length. The polynomial fit has ten adjustable parameters, $\nu$ and $W_c$ being fixed to their known values $\nu=1.6$ and $W_c=16.53$ (see \cite{romer}). In both cases the goodness of fit is quite acceptable (larger than 0.1). We find $y\approx 1.8$ for the unperturbed Anderson model and $y\approx 1.5$ for the rotated one. This compares well with the simulations of \cite{romer}.
}
\label{fig:Transi-tau2tildetyp-chb-L20-120}
\end{figure}

Note however that different system sizes $L$ lead to a different dependence of $D_2$ as a function of $t/\tau_\text{Th}$.
When comparing different system sizes, one should consider the Thouless time associated with the 3D Anderson model $T_\text{Th}=L^2/D$ (with $D$ the diffusion constant of the quasiperiodic kicked rotor) instead of $\tau_\text{Th}$ associated with the 1D quasiperiodic kicked rotor. This is done in Fig.~\ref{fig:Dqvstovtthouless-L20-120}. Then, the data for $D_2$ are seen to collapse onto a single curve, apart from deviations at small $t/T_\text{Th}$. This is at the expense of the physical meaning of the ergodic limit $D_q=d=3$ which in this latter case arises when $t/T_\text{Th}\approx 10^8$. This is due to the fact that we have rotated the eigenstates of the 3D Anderson model using a 1D diffusive dynamics. This implies a very strong anisotropy in 3D as the distance between two sites adjacent along the y-axis is $L$ in our vector of size $N$, and $L^2$ for those adjacent along the $z$ axis. It is then clear that the ergodic limit is reached only when the $z$-axis is filled, and this arises when $t$ reaches $\tau_\text{Th}=N^2/D$ where $N=L^3$ is the effective size of the sample along the $z$-axis.

The remarkable scale invariance observed in Fig.~\ref{fig:IPRvslboxovL} suggests that the rotated Anderson model remains critical at $W=W_c$, whatever the amplitude $t/T_\text{Th}$ of the perturbation. Indeed, away from criticality one expects to observe the emergence of one of the characteristic lengths discussed in IV A (the localization length or the correlation length). We have checked that the rotated eigenstates perform a localization-delocalization transition at the same value of disorder strength $W_c$ as the unperturbed ones. Following the analysis of \cite{romer}, we have considered the quantity $\tau_2 \equiv \ln P_2^\text{typ}/\ln \lambda$ with fixed $\lambda\equiv\ell/L=0.1$ where $P_2^\text{typ}=\exp(\langle \ln P_2\rangle)$ and $\langle . \rangle$ stands for an average over disorder realizations, see also Sec.~\ref{num_methods}. Figure \ref{fig:Transi-tau2tildetyp-chb-L20-120} shows the result of a finite-size scaling analysis for the standard Anderson model and the rotated one. In both cases, we find that our numerical data are {\it compatible} with a scaling 
$\tau_2 = F(L/\xi)+L^{-y} F_{\textrm{irr}}(L/\xi)$
where $F$ and $F_{\textrm{irr}}$ are scaling functions and
with the localization/correlation length $\xi\sim\vert W-W_c \vert^{-\nu}$ diverging at $W_c \approx 16.53$ with the critical exponent $\nu\approx 1.6$. The irrelevant exponent $y$ controls the usual irrelevant corrections (precise values are in the figure caption of Fig.~\ref{fig:Transi-tau2tildetyp-chb-L20-120} and quite compatible with the known results of \cite{romer}). The effect of the rotation is most clearly observed when considering the values of $\tau_2^c\equiv \text{lim}_{L\rightarrow \infty} \tau_2(W_c)$ extracted from the finite-size scaling analysis: $\tau_2^c \approx 1.66$ in case (a) and $\tau_2^c \approx 2.33$ in case (b) of Fig.~\ref{fig:Transi-tau2tildetyp-chb-L20-120}. The increase of $\tau_2^c$ with the perturbation strength $t/T_\text{Th}$ is in good agreement with the results of Fig.~\ref{fig:Dqvstovtthouless-L20-120}.

\subsection{Presence of a characteristic length }

\subsubsection{Intermediate map}

\begin{figure}[h]
    \begin{center}
     \includegraphics*[width=0.485\textwidth]{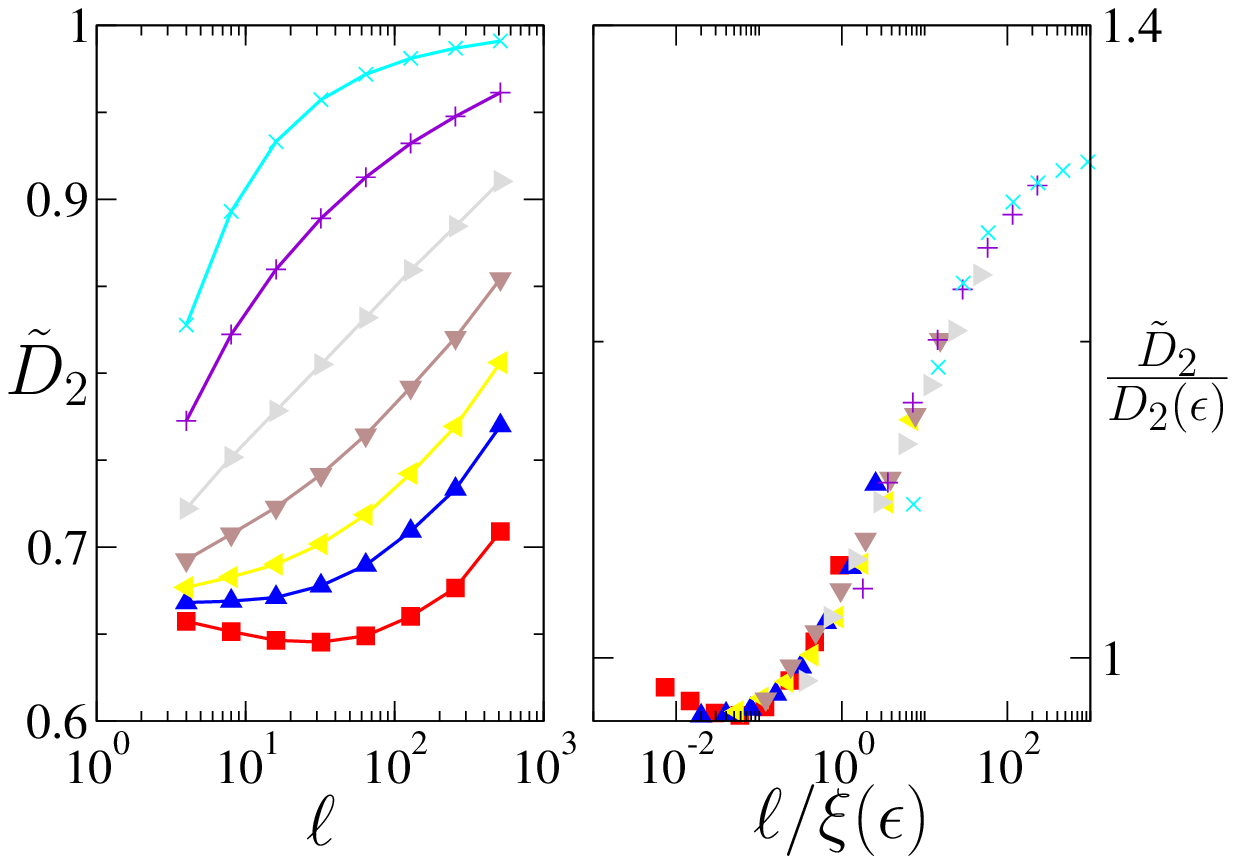}   \\
     \vspace{.3cm}
       \includegraphics*[width=0.485\textwidth]{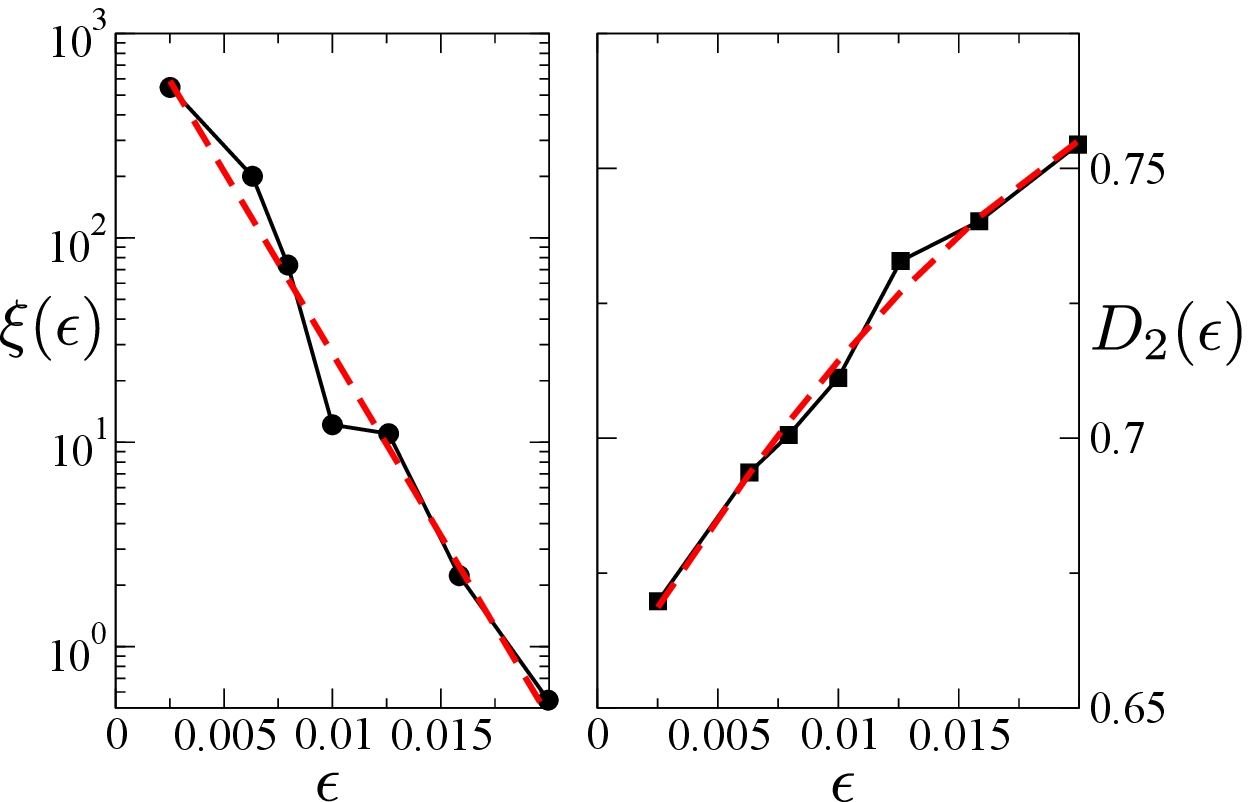}    
      \end{center}
        \caption{(Color online) Top: Local multifractal exponent for the random intermediate map, $\gamma=1/5$, $N=2^{13}$, $q=2$ under a generic change of basis. Left: raw data. Right: rescaled data. The points corresponding to $\ell=N/2$, $N/4$ and $N/8$ have been dropped. Each curve corresponds to a different value of the parameter $\epsilon$ in (\ref{deformId}). It is chosen to be of the form $\epsilon=10^{0.1 n-3}$; red squares: $n=4$; blue up triangles: $n=8$; yellow left triangles: $n=9$; brown down triangles: $n=10$; gray right triangles: $n=11$; purple pluses: $n=12$; cyan crosses: $n=13$.
Bottom: Left: Variation of the scaling length $\xi$ as a function of $\epsilon$. Right: variation of the second scaling parameter $D_2(\epsilon)$ as a function of $\epsilon$.  Black full symbols are exact data, dashed lines are fitting functions (resp. exponential and second order polynomial).
}
  \label{Dq_deformIdbis}
\end{figure} 

We now consider a change of basis of the random intermediate map (\ref{defU}). As for the PRBM model, we construct a deformation matrix (\ref{deformId}) and use it to transform the propagator $U$  into a new matrix $U'$.
 Results are displayed in Fig.~\ref{Dq_deformIdbis}, where  the local multifractal dimension $\tilde{D}_2$ is plotted as a function of the box size for various rotation parameters $\epsilon$. One clearly observes a systematic dependence on $\ell$ which hints to the presence of a characteristic length. However, contrary to the results in Sec. III, a rescaling of the boxsize is not sufficient to collapse the data. Indeed,  a {\it double} rescaling (both horizontal and vertical) is needed to collapse the data for
different $\epsilon$ values (see Fig.~\ref{Dq_deformIdbis} top). This indicates that there is a scaling length $\xi(\epsilon)$ for the box size 
and a scaling parameter for the multifractal dimension $D_2(\epsilon)$, both depending on $\epsilon$. Above the characteristic length, multifractality is destroyed (all multifractal dimensions equal to 1). Below 
the characteristic length, multifractality survives but the multifractal dimensions smoothly go to 1 when the perturbation is increased (see e.~g.~Fig.~\ref{Dq_deformIdbis} bottom right). The data are therefore fully compatible with a variant of scenario II including the presence of a characteristic length, below which multifractality is uniformly destroyed. Here the characteristic length does not come from the perturbation as in Section III A, but originates from the intrinsic characteristic length of the intermediate map $\Xi$ (see Section II) which is modified by the perturbation. 

\subsubsection{From momentum to position basis in the intermediate map}

\begin{figure}[h]
\begin{center}
\includegraphics*[width=0.4\textwidth]{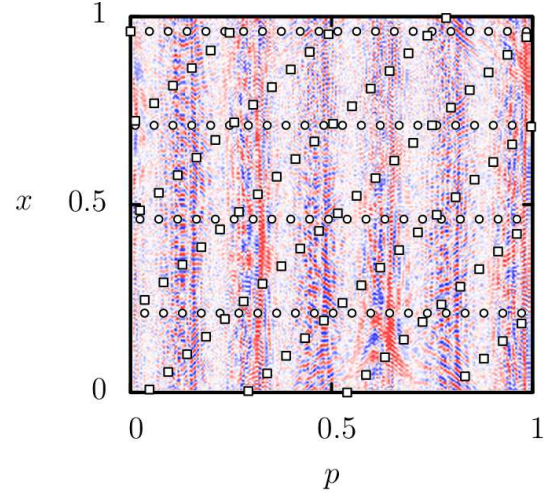}
\caption{(Color online) Discrete Wigner function of an eigenstate of the random intermediate map for $\gamma=1/3$, color/grayness varies from blue/dark gray (minimal value) to red/light gray (maximal value).
The symbols represent lines $a_1 P-a_2X=a_3$ (mod $2N$) ($x=X/2N$ and $p=P/2N$). Here $N=128$, $a_2=1$ and 
$a_1=64$ (circles) and $65$ (squares).
\label{lines}}
\end{center}
\end{figure}

\begin{figure}[h]
\begin{center}
\includegraphics*[width=0.45\textwidth]{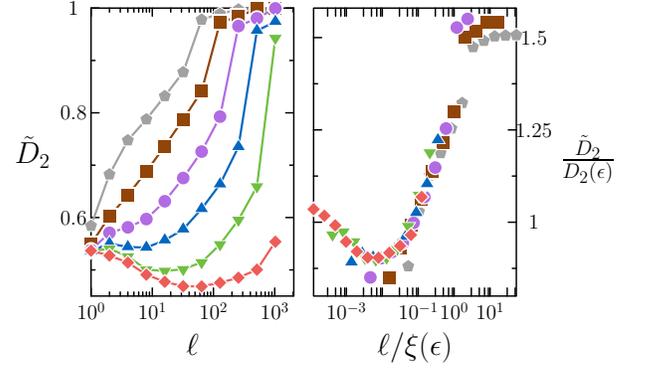}
\caption{(Color online) Left: $\tilde{D}_2$ as a function of the  box size $\ell$  for the random intermediate map with momentum to position basis change for $\gamma=1/3$ and different values of $N/a_1=2^5$ (pentagons), $N/a_1=2^4$ (squares), $N/a_1=2^3$ (circles),  $N/a_1=2^2$
 (triangles), $N/a_1=2^1$ (down-triangles), $N/a_1=1$ (diamonds), with $N=2^{13}$ (size of the Wigner
function is $2N$). Right: Rescaled  $\tilde{D}_2$ as a function of the rescaled
box size for the same data.
\label{Dqbases}}
\end{center}
\end{figure}

The intermediate map displays multifractal eigenvectors in the momentum representation \eqref{defU}. However in the position basis eigenvectors are extended.
It is natural to ask how multifractality is destroyed when one goes from one basis to the other. In this subsection we focus on a specific class of basis changes which interpolate between momentum to position basis and are linked to certain types of physical observables.

The Wigner function of a wavefunction $\psi$, in 
our case discrete Wigner function (DWF), is a quasi-probability distribution in phase space and 
thus provides an adequate testing ground to probe the transition from momentum to position. We use the DWF as described in \cite{miquel}. If the Hilbert 
space dimension is $N$ we define a phase-space grid of $2N\times 2N$ points. Let us label the 
$(2N)^2$ points $(X,P)$, with $1\leq X,P\leq 2N$. If the state of the system is given by a density matrix $\rho$, then the simplest expression for the DWF is
\begin{equation}
W(X,P)={\rm Tr}[\hat{A}(X,P)\rho]
\end{equation}
where $\hat{A}(X,P)$ are (so-called) point operators defined as the discrete Fourier transforms of the translation operators 
\begin{equation}
\label{transT}
T(a_1,a_2)=U^{a_1}V^{a_2}\exp[i\pi a_1 a_2/N],
\end{equation}
with $a_1,a_2$ integers. The translations in \eqref{transT} are defined by shifts $U$ and $V$ in position and momentum, such that $U\ket{X}=\ket{X+1}$ and $V\ket{P}=\ket{P+1}$. In \cite{miquel} it is shown that the DWF thus defined complies with all the properties expected from a Wigner function. Namely, $\sum_PW(X,P)$ is the  probability $|\psi_X|^2$ associated with the wavefunction in position representation. Similarly $\sum_X W(X,P)$ is associated with intensities in momentum representation. More generally, summing $W(X,P)$ along straight lines
\begin{equation}
\label{eq:line}
a_1 P- a_2 X= a_3\ \ ({\rm mod}\ 2N)
\end{equation}
with fixed $a_1,a_2$ and varying $a_3$ yields the probability distribution associated with the wavefunction expressed in the basis $\ket{a_3}$ of eigenvectors of $T(a_1,a_2)$. In particular for $a_1=1$ and $a_2=0$ we sum over vertical lines and we get momentum basis, while for $a_1=0$ and $a_2=1$ we get the position basis. Note that since  $N$ is a power of 2, whenever $a_1/a_2$ is also a power of 2 the lines will be horizontal for $a_1>a_2$ and vertical for $a_1<a_2$.  We illustrate this in Fig.~\ref{lines}. By changing $(a_1,a_2)$ we can go from position to momentum basis.

In Fig.~\ref{lines} we show an example of the (absolute value) of the DWF for one eigenstate of the random intermediate map with $\gamma=1/3$ as well as two examples of lines \eqref{eq:line}.

\begin{figure}[b]
\begin{center}
\includegraphics*[width=0.45\textwidth]{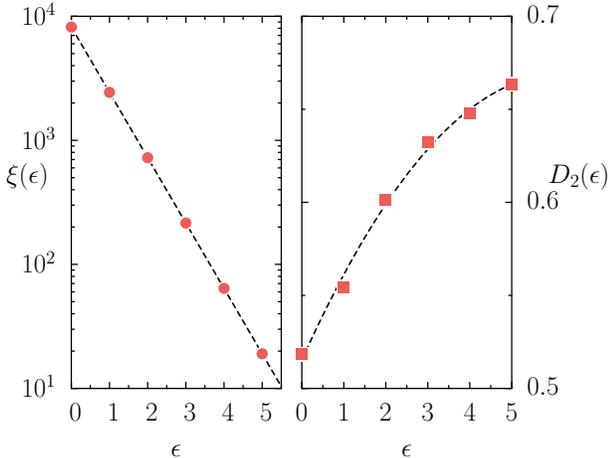}
\caption{(Color online) 
  Left: scaling length $\xi$ as a function of $\epsilon$ for the random intermediate map with momentum to position basis change for $\gamma=1/3$ and $N=2^{13}$. Right: second scaling parameter $D_2(\epsilon)$ as a function of $\epsilon=\log_2(N/a_1)$.  The dashed lines are fitting functions (resp. exponential and second order polynomial). Same data as in Fig.~\ref{Dqbases}.
\label{Dqbasesbis}}
\end{center}
\end{figure}

In Fig.~\ref{Dqbases} we show the local multifractal dimension $\tilde{D}_2(\ell)$ for different values of the slope $a_2/a_1$ of the lines defined in Eq.~(\ref{eq:line}) with $a_2=1$. In this case, the parameter $\epsilon=\log_2(N/a_1)$ gives the amplitude of the perturbation. The data displayed in Fig.~\ref{Dqbases} show that, as in the preceding case, a double rescaling enables to collapse the curves for different $\epsilon$. Again, a characteristic length $\xi(\epsilon)$ depending on $\epsilon$ separates two regimes. Above the scale $\xi(\epsilon)$, multifractality disappears, while below this scale it is uniformly and smoothly destroyed when $\epsilon$ increases (see Fig.~\ref{Dqbasesbis}). We conclude that for this more physical change of basis the multifractality breaks down following again a variant of the second scenario. 

\subsubsection{Change of basis with characteristic length scale in the Anderson transition}

\begin{figure}[b]
\begin{center}
\includegraphics*[width=0.45\textwidth]{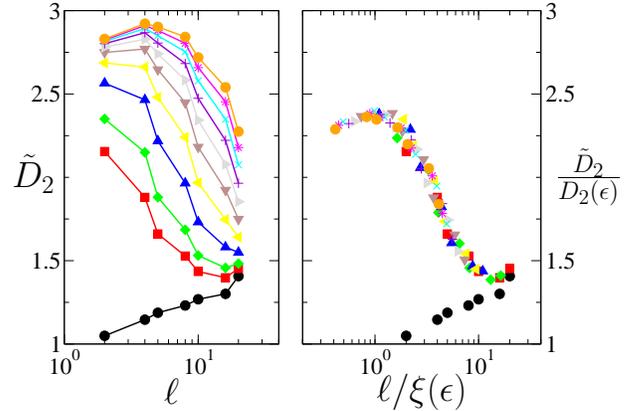}
\caption{(Color online) Double scaling analysis of the local multifractal dimensions for the Anderson model with isotropic basis change.
Left: $\tilde{D}_2$ as a function of the box size $\ell$
for different values of $\epsilon \equiv t/T_\text{Th}$: from bottom to top, $\epsilon=0$, $0.000125$, $0.00025$, $0.00062$,
$0.0013$, $0.0023$, $0.0035$, $0.005$, $0.0068$, $0.0089$, $0.011$. Here $T_\text{Th}=L^2/D$, $L=80$ and $D\approx 0.403$ the diffusion
constant determined numerically for the 3D isotropic periodic Kicked Rotor \eqref{eq:H3DKR} with $K=10$ and $\hbar=2.89$. Right:
Rescaled $\tilde{D}_2$ as a function of the rescaled
box size for the same values of $\epsilon$.
\label{And3D1}}
\end{center}
\end{figure}

\begin{figure}[t]
\begin{center}
\includegraphics*[width=0.45\textwidth]{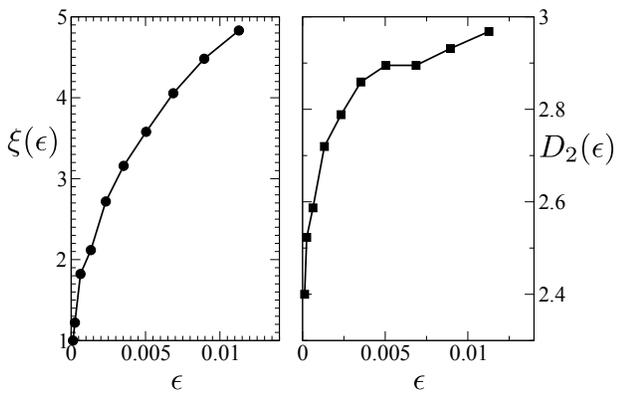}
\caption{ Left: box size scaling $\xi(\epsilon)$ as a function of $\epsilon$ for the Anderson model with isotropic basis change. Right: scaling parameter
$D_2(\epsilon)$ as a function of $\epsilon$. Same data as in Fig. 26.
\label{And3D2}}
\end{center}
\end{figure}

In the two previous cases considered, a length scale appears below which multifractality is smoothly destroyed. In the case of the PRBM
model however, there is no such feature. We believe that the difference reflects the fact that the intermediate map has an intrinsic length scale
$\Xi=N/b$ while the PRBM model does not. The change of basis rescales this characteristic length and that is the reason for scaling
laws with two parameters.

In this respect, the Anderson model is similar to the PRBM model: it has no intrinsic length scale in the critical regime. The
previous basis change using the quasiperiodic kicked rotor has a characteristic length $\Lambda=\sqrt{D t}$ associated with diffusion, but 
since $\Lambda\gg L$ for the parameters considered so far, this characteristic length does not play any role, as observed in Fig.~\ref{fig:IPRvslboxovL}. 

In order to study the effect of such a length scale on the basis change, we have to work in a regime where $\Lambda <L$. Therefore we consider a rotation with the evolution operator of a 3D kicked rotor:
\begin{equation}\label{eq:H3DKR}
 H_{\mathrm{3KR}} = \frac{\boldsymbol{p}^2}{2}+ K \frac{\cos x+\cos y + \cos z}{3} \sum_n \delta(t-n) \; ,
\end{equation}
with $K=10$ and $\hbar=2.89$. For this choice of parameters, the 3D kicked rotor (61) is in the delocalized phase and displays an isotropic 3D diffusion.  We considered the evolution over
a certain number of periods $t$ of the critical states of the Anderson model. We analyzed their multifractal properties by considering
the behavior of $\tilde{D}_2$ as a function of the box size $\ell$, as plotted in Fig.~\ref{And3D1}. The curves obtained at different
perturbation strengths $\epsilon=t/T_\text{Th}$ depend systematically on $\ell$ and $\epsilon$ and collapse onto a single scaling curve when $\tilde{D}_2$ is rescaled by $D_2(\epsilon)$ and $\ell$ is rescaled by $\xi(\epsilon)$. This strongly suggests a
picture similar to the scenario obeyed by the intermediate map under a basis change: at small scales $\ell \ll \xi(\epsilon)$,
multifractality is smoothly and uniformly changed by the perturbation, with $D_2(\epsilon)$ going from its unperturbed value at $\epsilon=0$, $D_2(0)\approx 1.3$ to $D_2(\epsilon)=3$ when $\epsilon\rightarrow 1$ (see Fig.~\ref{And3D2}). 

However, in the present case, $\xi(\epsilon)$ varies similarly to $\Lambda=\sqrt{D t}$, thus increases as a function of
$\epsilon$. In addition, at large scales $\ell\gg\xi( \epsilon)$, the unperturbed multifractality is recovered. In a sense, the
multifractal critical states are coarse grained to a size $ \Lambda $ by diffusive evolution, which affects multifractality only at
small scales $\ell \ll \Lambda$.

\section{Conclusion}
\label{conclusion}

In this paper, we have studied the destruction of quantum multifractality in the presence of different natural perturbations. The models we considered are representative of several classes of systems displaying quantum multifractality. The perturbations have been chosen to represent potential experimental constraints in realistic systems. Our numerical and analytical results confirm the conjecture presented in \cite{short}. We found that multifractality can be destroyed in the presence of a perturbation following two scenarios. In scenario I, the perturbation introduces a characteristic scale below which multifractality is unchanged, and above which it is completely destroyed. In our case, this describes the smoothing of the singularity in the intermediate map (Section III) and the Anderson model away from the critical point (Section IV A). Scenario II corresponds to a uniform destruction of multifractality at sufficiently small scales. Depending on the presence of a characteristic scale in the system, there can be two variants of scenario II. If there is no characteristic scale, multifractality is the same at all scales and smoothly goes to the ergodic value for large perturbation. This is illustrated by the case of the intermediate map for a change of slope (Section IV B), and by the change of basis for the PRBM model (Section V A 1) and the Anderson model (perturbation without characteristic scale, Section V A 2). If the system has a characteristic scale, scenario II corresponds to the uniform destruction of multifractality as before but only below this characteristic scale. This behavior can be revealed by a double scaling analysis. Above the characteristic scale, multifractality can be completely absent, as in the case of the change of basis in the intermediate map (Sections V B 1 and 2), or similar to the unperturbed system as in the case of the Anderson model when the change of basis has a characteristic length (Section V B 3).  Thus the image presented in \cite{short} is confirmed in this more detailed analysis, but our results show that subtle variations on this broad picture can appear.

The results presented in this paper also give some insight concerning the experimental observation of multifractality in various systems. Indeed, experimental setups are unavoidably subject to imperfections, which will act as perturbations of the ideal model that is implemented, including the measurement in a non optimal basis. The results of Section III show that smoothing the singularity of the kicked potential in the intermediate map preserves the original multifractality below a certain scale, which imposes a minimal resolution to the experimental measurements. This kind of perturbation can appear e.g. if the model is implemented with photonic crystals. The truncation of a Fourier series to simulate the intermediate map could be envisioned in a cold atom context, but here our results show that a huge number of harmonics are needed, and other techniques should be devised to implement the singular potential. Section IV shows that one can afford an imprecision on the slope of the potential for finite-size systems, but multifractality will be modified, however in a way which can be precisely predicted. 
In an experiment, there will be more natural observables corresponding to specific measurement bases. We have investigated the behavior of multifractal properties under a change of basis, by interpolating between the momentum and position bases, or using a generic change of basis built from random unitary matrices.  It turns out that a modified multifractality is observable for small rotations of the basis, but subtle behaviors can emerge depending on the presence or absence of a characteristic scale, originating either from the model or the perturbation. This is particularly striking in the case of the Anderson model, where different change of bases could lead to different variants of our second scenario. Interestingly, our results confirm that the change of basis in this case conserves the criticality of the model.

Despite many theoretical works in the recent past, direct experimental observation of multifractality on quantum wave functions has remained elusive up to date. Our results show that experimental imperfections will eventually destroy the multifractal properties if they exceed a certain level, but that a range of parameters subsists where multifractality could be observed.  The scenarios for multifractality breakdown confirmed in this paper could guide the design of future experiments, which would reliably detect multifractality for small imperfection strength and observe the scenarios for larger perturbations.

\acknowledgments We thank Olivier Herscovici and Claudio Castellani for discussions and insights. We thank CalMiP for access to its supercomputers and the University Paul Sabatier (OMASYC project). This work was supported by Programme Investissements d'Avenir under the program ANR-11-IDEX-0002-02, reference ANR-10-LABX-0037-NEXT, and by the ANR grant K-BEC No ANR-13-BS04-0001-01. JM is grateful to the University of Li\`{e}ge for the use of the NIC4 supercomputer (SEGI facility), and for funding (project C-13/86).
I.G.M. received support from ANCyPT grant PICT 2010-1556 and from CONICET grant PIP 114-20110100048. I.G.M. and J.M. received support from CONICET-FNRS binational project, and B.G. and I.G.M. from the CONICET-CNRS bilateral project PICS06303.

\end{document}